\documentclass[fleqn,usenatbib]{mnras}

\usepackage{newtxtext,newtxmath}

\usepackage[T1]{fontenc}

\DeclareRobustCommand{\VAN}[3]{#2}
\let\VANthebibliography\thebibliography
\def\thebibliography{\DeclareRobustCommand{\VAN}[3]{##3}\VANthebibliography}

\usepackage{graphicx}	
\usepackage{amsmath}	
\usepackage{soul}
\usepackage{mathtools}
\usepackage{amsfonts}
\usepackage{bbm}
\usepackage{subfigure}
\usepackage{xspace}
\usepackage{appendix}
\usepackage[belowskip=0pt,aboveskip=0pt]{caption}
\usepackage[capitalise]{cleveref}   
\usepackage{orcidlink}
\makeatletter
\usepackage{etoolbox}
\usepackage[normalem]{ulem}
\patchcmd\H@refstepcounter{\protected@edef}{\protected@xdef}{}{}
\makeatother

\crefname{subsection}{Section}{Sections}
\Crefname{section}{Section}{Sections}
\Crefname{equation}{equation}{equations}
\renewcommand{\vec}[1]{\boldsymbol{#1}}
\newcommand{\normal}[2]{\mathcal{N}(#1, #2)}
\newcommand{\xv}[0]{\vec{x}}
\newcommand{\yv}[0]{\vec{y}}
\newcommand{\Xv}[0]{\vec{X}}
\newcommand{\Yv}[0]{\vec{Y}}
\renewcommand{\hat}[1]{\widehat{#1}}

\newcommand{\cdm}{$\Lambda$CDM\xspace}
\newcommand{\rockstar}{\textsc{Rockstar}\xspace}
\newcommand{\consistrees}{\textsc{consistent-trees}\xspace}
\newcommand{\diffmah}{DiffMAH\xspace}
\newcommand{\multicam}{MultiCAM\xspace}
\newcommand{\camopt}{CAM $a(\mopt)$\xspace}

\newcommand{\mvir}{M_{\rm vir}}
\newcommand{\xoff}{x_{\rm off}}
\newcommand{\ToverU}{T / | U |}
\newcommand{\cvir}{c_{\rm vir}}
\newcommand{\munit}{h^{-1} M_{\odot}}
\newcommand{\aopt}{a_{\rm opt}}
\newcommand{\mopt}{m_{\rm opt}}

\newcommand{\spin}{\lambda_{\rm bullock}}
\newcommand{\vmaxovervvir}{V_{\rm max}/V_{\rm vir}}
\newcommand{\simsample}{\texttt{M12}\xspace}
\newcommand{\rhosp}{\rho_{\rm sp}}
\newcommand{\Mpeak}{M_{\rm peak}}
\newcommand{\tauc}{\tau_{c}}
\newcommand{\alphaearly}{\alpha_{\rm early}}
\newcommand{\alphalate}{\alpha_{\rm late}}
\newcommand{\tdyn}{t_{\rm dyn}}
\newcommand{\yhalo}{Y_{\rm halo}}
\newcommand{\xmah}{X_{\rm mah}}
\newcommand{\Msun}{{\rm M}_\odot}

\title[MultiCAM]{\multicam: a multivariable framework for connecting the mass accretion history of haloes with their properties}

\author[I. Mendoza et al.]{
Ismael Mendoza \orcidlink{0000-0002-6313-4597},$^{1}$\thanks{E-mail: imendoza@umich.edu}
Philip Mansfield \orcidlink{0000-0001-9863-5394},$^{2,3}$
Kuan Wang \orcidlink{0000-0001-7690-2260}$^{1,4}$
and Camille Avestruz \orcidlink{0000-0001-8868-0810}$^{1,4}$
\\
$^{1}$Department of Physics, The University of Michigan, Ann Arbor, MI 48109 USA\\
$^{2}$Kavli Institute of Particle Astrophysics and Cosmology, Stanford University, Stanford, CA 94305, USA\\
${^3}$SLAC National Accelerator Laboratory, Menlo Park, CA 94025, USA\\
$^{4}$Leinweber Center for Theoretical Physics, University of Michigan, 450 Church St, Ann Arbor, MI 48109, USA\\
}

\date{Accepted 2023 May 16. in original form 2023 February 15}

\pubyear{2023}

\begin{document}
\label{firstpage}
\pagerange{\pageref{firstpage}--\pageref{lastpage}}
\maketitle

% Abstract of the paper
\begin{abstract}
Models that connect galaxy and halo properties often summarize a halo's mass accretion history (MAH) with a single value, and use this value as the basis for predictions. However, a single-value summary fails to capture the complexity of MAHs and information can be lost in the process.
We present \textit{\multicam}, a generalization of traditional abundance matching frameworks, which can simultaneously connect the full MAH of a halo with multiple halo and/or galaxy properties.
As a first case study, we apply \multicam to the problem of connecting dark matter halo properties to their MAHs in the context of a dark matter-only simulation. While some halo properties, such as concentration, are more strongly correlated to the early-time mass growth of a halo, others, like the virial ratio, have stronger correlations with late-time mass growth. This highlights the necessity of considering the impact of the entire MAH on halo properties. 
For most of the halo properties we consider, we find that \multicam models that use the full MAH achieve higher accuracy than conditional abundance matching models that use a single epoch.
We also demonstrate an extension of \multicam that captures the covariance between predicted halo properties. This extension provides a baseline model for applications where the covariance between predicted properties is important.
\end{abstract}


%%%%%%%%%%%%%%%%%%%%%%%%%%%%%%%%%%%%%%%%%%
\section{Introduction}\label{sec:intro}
%%%%%%%%%%%%%%%%%%%%%%%%%%%%%%%%%%%%%%%%%%

% 1. Introduce DM haloes and connection to galaxies.
Characterizing the properties and growth of dark matter haloes has been an important goal of cosmological \textit{N-body} simulations \citep{diemand2011structure,frenk2012dark}. % Two big review papers on dark matter halo, CDM simulations, and LSS.
Dark matter haloes are groups of dark matter particles that have gravitationally collapsed into bound structures. In the \cdm cosmological model, every galaxy forms within the potential well provided by a dark matter halo \citep{white1978core, blumenthal1984formation}. 
Thus, galaxies and their dark matter haloes are closely connected, meaning that models which attempt to predict the properties of galaxies must account for the behaviour and properties of their dark matter haloes \citep[e.g.][]{hearin2013darkside,hearin2016introducing,wechsler2018connection} % CAM paper, decorated HOD paper, and review on galaxy-halo connection.

% 2. Properties of halo and mass growth are strongly correlated
Previous work has established a deep connection between a halo's present-day ($z=0$) properties and its \textit{mass accretion history (MAH)}, i.e. its mass growth as a function of time. Properties such as concentration, virial ratio, centre of mass offset, spin, axis ratio, and splashback radius have been studied in relation to MAH. Early-forming haloes tend to have a higher concentration on average than late-forming haloes \citep[e.g.][]{wechsler2002concentrationsfromassembly}, and merger events induce lasting changes in halo structure which are encoded as a universal signatures in the halo's concentration \citep[e.g.][]{wang2020concentrations}. Other properties like the centre of mass offset, virial ratio, and splashback radius have strong correlations with the halo's recent mass growth history and merging activity \citep[e.g.][]{power2012dynamical,tae2023splashback}. This joint dependence leads to substantial covariance between halo parameters \citep[e.g.][]{lau2021correlations}. Much of this dependence comes from long-term growth trends: it has been found that a significant percentage of the variance in the concentration, axis ratio, and spin of a dark matter halo can be explained by the first principal component of the mass assembly history \citep[e.g.][]{chen2020relating}.

% 3. Specific instance demonstrating importance of connection between halo properties and growth: dynamical state and relaxedness.
The MAH of a halo directly impacts the dynamical state of a halo, which in turn determines the reliability of structural measurements of its properties. Previous studies have established that haloes that have recently experienced one or more major mergers are more likely out of dynamical equilibrium \citep{tormen1997structure, hetznecker2006evolution}. These major merger events can cause temporary deviations from a halo's equilibrium state during which its structural properties change rapidly and might not be well defined \citep{ludlow2016mass}. Thus, it is critical that we characterize the dynamical state of haloes so that their structural measurements can be robustly propagated to downstream analysis.
Previous work measuring the distribution of halo properties in simulations attempted to address this by selecting a subsample of \textit{relaxed haloes}, i.e., those haloes considered to be close to dynamical equilibrium \citep[e.g.][]{neto2007statistics,klypin2011dark,klypin2016multidark}. A closely related line of work seeks to identify relaxed galaxy clusters to avoid similar biases in the corresponding measurements \citep[e.g.][]{cui2017dynamical,zhang2022three}. However, there is a significant ambiguity on how to exactly define this relaxed sample for both cases, which usually rely on hard-cuts. This further highlights the need for increasing our understanding of the relationships between a galaxy's or halo's properties, MAH, and dynamical state. 

% 4. Transition to methods that connect growth and properties, mention CAM and limitations.
A common way to connect galaxy or halo properties to their MAH is to use a single parameter summary of the MAH, such as the half-mass scale \cite[e.g.][]{gao2005age} or the value returned by a single-parameter fit \citep[e.g.][]{wechsler2002concentrationsfromassembly}. This framework leads to a one-to-one parameter correlation analysis called abundance matching, which corresponds to a prediction model that {\it assumes perfect correlation} between the two parameters (e.g. half-mass scale and halo concentration) \citep{kravtsov2004dark}. Abundance matching and its hierarchical extension, conditional abundance matching \citep[CAM,][see \cref{sec:cam} for a description of these methods]{hearin2014dark}, have been effective models for a range of applications. For example, CAM can predict low-redshift galaxy statistics like two-point correlation functions in SDSS to reasonable accuracy \citep{hearin2014dark}. However, the MAH of a dark matter halo is a complex multidimensional quantity that contains richer predictive information than single parameter summaries. 

\begin{table}
\centering
\begin{tabular}{|c|c|}
\hline
Parameter & Value \\ [0.5ex]
\hline\hline
Box size & 250 $\text{Mpc} \, h^{-1}$ \\ \hline
Number of particles & $2048^{3}$ \\ \hline
Particle mass & $1.35 \times 10^{8}\, \Msun h^{-1}$ \\ \hline
Force resolution & $1.0\, \text{kpc} \, h^{-1}$ \\ \hline
Initial redshift & $80$ \\ \hline 
Number of snapshots & $180$ \\ \hline 
Hubble parameter $h$ & $0.7$ \\ \hline
$\Omega_{\Lambda}$ & $0.73$ \\ \hline
$\Omega_{m}$ & $0.27$ \\ \hline
$\Omega_{b}$ & $0.0469$ \\ \hline
Tilt $n$ & $0.95$ \\ \hline
$\sigma_{8}$ & $0.82$ \\ \hline
\end{tabular}
\caption{
    Simulation and cosmological parameters of the Bolshoi dark matter-only cosmological \cdm simulation presented in \protect \cite{klypin2011dark}, which is based on the WMAP5 cosmology \citep{dunkley2009five}.
}
\label{tab:simulation-params}
\end{table}

% 5. Mass accretion history is complex
MAHs are typically made up of a smooth accretion component consisting of an early-fast accretion phase and a late-slow accretion phase, which was successfully captured with a three-parameter model in \citep{hearin2021differentiable}. The MAH also includes a non-smooth accretion component in the form of an arbitrary number of discrete major merger events that can significantly change halo properties on a short time-scale \citep[e.g.][]{hetznecker2006evolution,power2012dynamical,wang2020concentrations}. Separately, it has been shown that different present-day halo properties correlate more or less strongly with different parts of the MAH \citep[e.g.][]{wong2012dark}. Thus, summarizing the MAH with a single quantity leads to discarding a significant amount of useful information. Another significant drawback of one-to-one parameter models is that they are unable to capture the covariance between predictions. If the same single parameter MAH summary is chosen, CAM-like models necessarily output a perfect correlation between any pair of predicted halo properties. Thus, if one is interested in emulating multiple halo properties from a given MAH, one-to-one models are insufficient. 

%6. Brief MultiCAM intro
To address the aforementioned limitations we propose a new method for connecting galaxy or halo properties with their formation history: \textit{\multicam}. \multicam is a generalization of the traditional abundance matching framework that consistently incorporates the full formation history into a prediction of single-epoch properties while preserving the key benefits of CAM. \multicam utilizes the full covariance between features and targets in its predictions. Moreover, \multicam can predict multiple properties simultaneously and correctly capture the correlations between them.
As a first demonstration of our new method, we apply it to connecting dark matter halo properties with their MAH. In the future, our main focus will be in applying this method to predict baryonic properties.

% 7. Paper organization
This paper is organized as follows. \cref{sec:datasets} describes the simulation suite and halo sample used in our studies. \cref{sec:methods} presents the parametrizations of MAH we consider in this work, gives an overview of CAM, and a detailed description of \multicam. In \cref{sec:results} we characterize the covariance of MAH and halo present-day properties, and evaluate \multicam on our halo sample. \cref{sec:discussion} discusses future applications of \multicam and how it compares to other methods. Finally, in \cref{sec:conclusions} we present our conclusions.

%%%%%%%%%%%%%%%%%%%%%%%%%%%%%%%%%%%%%%%%%%
\section{Data set}\label{sec:datasets}
%%%%%%%%%%%%%%%%%%%%%%%%%%%%%%%%%%%%%%%%%%

%==============================
\subsection{Simulation suite} \label{sec:sim-suite}
%==============================

For our data set we use the Bolshoi dark matter-only cosmological simulation \citep{klypin2011dark} which was performed with the Adaptive-Refinement-Tree (ART) code described in \citep{kravtsov1997adaptive}. The simulation has outputs at $180$ snapshots starting at $a_{179}=0.07835$ and ending at $a_{0}=1.00035 \approx 1$. The spacing between early snapshots is $\Delta a = 0.006$ between $a_{179} = 0.07835$ and $a_{77}= 0.80835$, and $\Delta a = 0.003$ between late snapshots $a_{77} = 0.80835$ and $a_{0}= 1.00035$. The cosmological parameters and other simulation details are shown in \cref{tab:simulation-params}.

The halo catalogues were generated by the \rockstar halo finder \citep{behroozi2012rockstar}, as run by \citet{rodriguez_peubla2016demographics}. This catalogue uses both position and velocity information to identify each halo in the simulation.
Halo finder comparison projects have found this algorithm to perform well at halo finding tasks, including detecting substructure and tracing mergers \citep[e.g.,][]{knebe2011halosgonemad}.

We use catalogues generated by \textsc{consistent-trees} \citep{behroozi2012gravitationally} to construct the merger history that we use for our analysis \citep{rodriguez_peubla2016demographics}. 
Given a merger event, we define the \textit{main progenitor} halo as the one that contains the most particles that end up in the resulting halo after the merger. 
Given a present-day ($z=0$) halo, a \textit{merger tree} can be constructed by following its evolution at each snapshot in the simulation going backwards in time. The \textit{main progenitor branch} of a given present-day halo is the branch in the merger tree resulting from following the main progenitor halo backwards in time at each snapshot. 

%======================================================
\subsection{Defining the halo sample}
%======================================================
Throughout this work we use the same data set of a random sample of $10,000$ haloes from the Bolshoi simulation in the mass bin of $M_{\rm vir}\in[10^{12},\, 10^{12.2}]\munit$ which we denote as \simsample. Here, $M_{\rm vir}$ is the bound mass within a radius enclosing an average density corresponding to the overdensity threshold defined in \citet{bryan1998xrayclusters}. We take this radius to be the virial radius $R_{\rm vir}$.

For each of the haloes in this sample, we use the \rockstar catalogue at each snapshot and \consistrees to extract the corresponding main progenitor branch and the virial masses of progenitors at each snapshot in this branch. We do not use all of the $180$ snapshots in the Bolshoi simulation, rather we impose a cut-off based on the mass resolution of our simulation. We pick our first snapshot to be the earliest snapshot out of the $180$ where at most $5\%$ of haloes have a virial mass lower than $50$ times the particle mass. This ensures that we never attempt to analyse snapshots where a substantial portion of our sample is unresolved. For our \simsample sample, we consider a total of $N_{\rm snap} = 165$ scales ranging from $a_{164} = 0.18635$ up to $a_{0} = 1$. 

For the small percentage $\leq 1 \%$ of haloes in our sample that do not have a corresponding main line progenitor at $a_{164}$ (or in any subsequent snapshots), we assign them a virial mass at those missing snapshots to be the mass of a single particle of the simulation. This is so that there are no missing values in the MAHs for all haloes in our \simsample sample. 

%==============================
\subsection{Halo properties and their convergence} \label{sec:halo-properties}
%==============================

In this study we mainly consider halo concentration, $\cvir,$ defined as the ratio of the virial radius to the NFW scale radius; the normalized maximum value of the halo's rotation curve $\vmaxovervvir;$ the offset between the halo's centre of mass and its most bound particle $\xoff;$ the virial ratio, $\ToverU;$ its dimensionless spin parameter, $\lambda_{\rm bullock};$ and its second minor-to-major axis ratio, $c/a$. See \citet{mansfield2021biased} for the exact definitions of these properties as computed by \rockstar. 

\citet{mansfield2021biased} measured the minimum converged masses for each of these properties in Bolshoi at different levels of acceptable numerical bias. No detectable bias is measured in $\vmaxovervvir$ at $M_{\rm vir} > 10^{11.8}\,\munit$ at $M_{\rm vir} > 10^{11.6}\,\munit,$ $T/|U|$ at $M_{\rm vir} > 10^{11.1}\,\munit,$ $\lambda_{\rm bullock}$ at $M_{\rm vir} > 10^{10.2}\,\munit,$ and $c/a$ at $M_{\rm vir} > 10^{10.9}\,\munit.$ \citet{mansfield2021biased} do not report a $c_{\rm vir}$ convergence limit for Bolshoi, but do report a $c_{\rm vir}$ convergence limit for Erebos\_CBol\_L125 \citep{diemer2015universal} at $M_{\rm vir} > 10^{11.6}\,\munit,$ which has an identical cosmology, identical particle mass, coarser force softening, and coarser time-steps than Bolshoi. Therefore, all the considered properties are converged within our mass window of $[10^{12},$ $10^{12.2}]\munit.$

We also briefly consider several other, more minor halo properties. For example, the average of the first minor-to-major axis ratio $b/a$ and the second minor-to-major ratio $c/a$, which we denote with $q$:
\[
    q = \frac{1}{2} \left( \frac{b}{a} + \frac{c}{a} \right).
\]
Because this property is derived from $b/a$ and $c/a$, it is converged at about $10^{10.9}h^{-1}M_\odot$. For all other halo properties, their definitions can be found in \citet{mansfield2021biased} and they are also converged within our mass window.

%%%%%%%%%%%%%%%%%%%%%%%%%%%%%%%%%%%%%%%%%%
\section{Methods}\label{sec:methods}
%%%%%%%%%%%%%%%%%%%%%%%%%%%%%%%%%%%%%%%%%%

\begin{figure*}
    \centering
    \includegraphics[width=\textwidth]{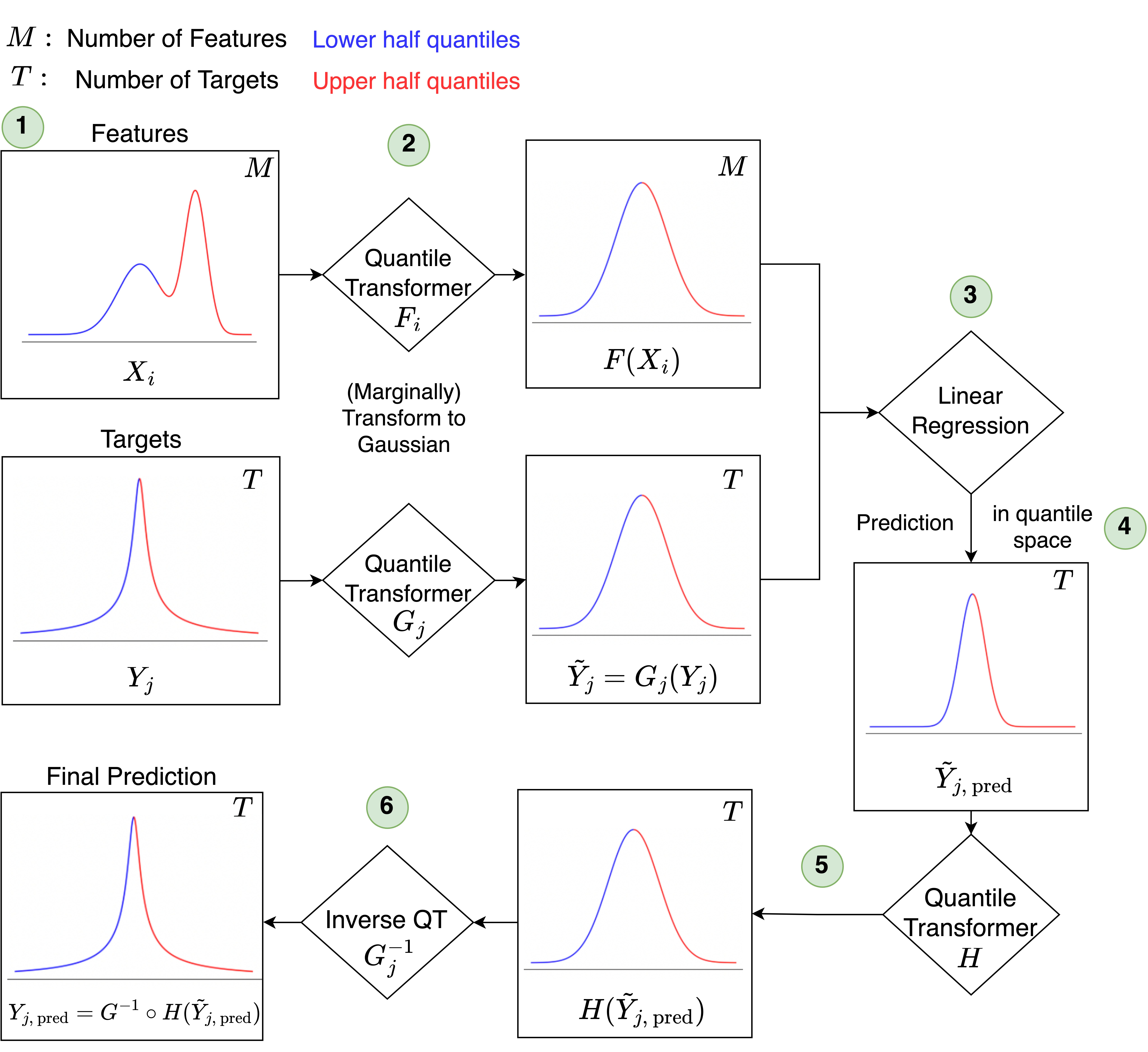}
    \caption{
        \textbf{Schematic illustrating the \multicam method.} In this diagram we illustrate the novel method presented in this work to connect MAH information to present-day halo properties: `\multicam'. 
        Each step of our algorithm is marked with a green circle. Each box represents the 1D distribution of one of the $M$ features or $T$ targets. The curve of each 1D distribution is delineated so that the blue and red curves intersect at the median. The rhombuses represent algorithms, either a quantile transformer to marginally map variables to Gaussian distributions, or a linear regression prediction model. The algorithm and each of the steps are described in detail in \cref{sec:multicam-general}.
    } 
\label{fig:cartoon}
\end{figure*}

%==============================
\subsection{Parametrizations of MAHs} \label{sec:mah-summaries}
%==============================

First, we introduce the notation that we use to parametrize the MAHs and their properties. We measure time through the cosmological scale factor:
\begin{equation}
    a(z) = \frac{1}{1+z}
\end{equation}
We track mass growth through the normalized peak mass, 
\begin{equation}
    m(a) = \frac{\Mpeak(a)}{\Mpeak(a = 1)},
    \label{eq:mass-fraction}
\end{equation}
where we take the ratio of $\Mpeak$ values to force monotonicity,
\begin{equation}
    \Mpeak (a) = \max_{0 \leq a' \leq a}[\mvir(a')].
\end{equation}
The difference between $\Mpeak(a)$ and $\mvir(a)$ is significant for subhaloes due to the large amount of mass loss they experience \citep[e.g.][]{wechsler2018connection}, but the difference is less important for the central haloes in \simsample, since their masses will typically increase over time. The main impact on our sample of host haloes is that it allows $m(a)$ to be inverted. To that end, we define
\begin{equation}
    a(m) = m(a)^{-1}.
\label{eq:am-ma-relationship}
\end{equation}
Since $m(a)$ is monotonic, but not strictly increasing, we take $a(m)$ to be the first scale factor at which the halo reaches a given mass. When inverting $m(a)$ we use piecewise power-law interpolation between adjacent snapshots of a halo's MAH.

In addition, per convention, we sometimes use the notation $a_{1/n}$ where $n$ is an integer to mean: 
\begin{equation}
    \label{eq:a_1/n}
    a_{1/n} = a(m = 1/n).
\end{equation}
This notation, usually with $n=2$, is often used in the literature as a tracer of formation \citep[e.g.][]{gao2005age}.

We define a halo's dynamical time $\tdyn$ as the time it takes for a test particle to travel a distance of virial radius $R_{\rm vir}$ at a speed of $V_{\rm vir}$, the orbital speed of a particle on a circular orbit at $R_{\rm vir}$. Since all haloes have the same enclosed density within $R_{\rm vir},$ $t_{\rm dyn}$ is only a function of redshift and cosmology:
\begin{align}
    t_{\rm dyn} &= \frac{R_{\rm vir}}{\sqrt{GM_{\rm vir}/R_{\rm vir}}} = \frac{1}{H(z)} \left(\frac{2\rho_c(z)}{\rho_{\rm vir}(z)}\right)^{1/2} \label{eq:tdyn} \\
    &= 2.01\,{\rm Gyr} \left(\frac{\rho_{\rm vir}(z)/\rho_{\rm c}(z)}{97.0}\right)^{-1/2} \left(\frac{H(z)}{70\ {\rm km\,s^{-1}\,Mpc^{-1}}}\right)^{-1}. \label{eq:tdyn_norm}
\end{align}
For convenience, \cref{eq:tdyn_norm} is normalized to the $z=0$ virial density in the Bolshoi simulation. Following from this definition, $m_{\tdyn}$ is the mass fraction at a time $\tdyn$ before the present day. $m_{\tdyn}$ is a commonly used measure of late-time accretion rates and its unnormalized equivalent is tracked by \consistrees catalogues by default.

We also analyse the best-fitting exponential scale factor of each MAH \citep{wechsler2002concentrationsfromassembly}:
\begin{equation}
    M_{\rm vir}(z)  / M_{\rm vir}(z = 0) = e^{-\alpha z}.
\label{eq:exponental_M(z)}
\end{equation}

%==============================
\subsection{\diffmah model of smooth MAHs} \label{sec:diffmah-definitions}
%==============================

We also consider the best-fitting parameters of the \diffmah model of smooth MAHs presented in \cite{hearin2021differentiable}. This model consists of the following fitting function: 
\begin{equation}
    \Mpeak (t) / \Mpeak(t=t_{0}) = (t/t_{0})^{\alpha(t)} \label{eq:diffmah-base}
\end{equation}
where $t$ is age of the universe, and $t_{0}$ is the present-day age of the universe. Finally, $\alpha(t)$ is a sigmoid function defined as: 
\begin{equation}
    \alpha(t; \tauc, k, \alphaearly, \alphalate) \equiv \alphaearly + \frac{\alphalate - \alphalate}{1 + \exp ( - k (t - \tauc))}
    \label{eq:diffmah-exponent}
\end{equation}
and has parameters $\alphaearly, \alphalate, k$, and $\tauc$ with an explicit physical meaning. First, $\alphaearly, \alphalate$ determine the asymptotic value of the power-law index at early and late times respectively; $\tauc$ controls the transition time between the early- and late-time indices; and $k$ determines the speed of transition between the two phases. As in \cite{hearin2021differentiable}, we fix $k=3.5$.

%======================================================
\subsection{Statistical algorithms} \label{sec:statistical-algorithms}
%======================================================

In this section we introduce the statistical algorithms we use for predictions connecting MAH and present-day halo properties. 

%---------------------------------------------------------
\subsubsection{Conditional abundance matching} \label{sec:cam}
%---------------------------------------------------------

% Introduce CAM with some context of where it is used
One of the methods we use is an adapted \textit{Conditional-Abundance Matching (CAM)}. The CAM algorithm is a method which was originally developed to study and model the connection between halo ages --- traced through properties like $a_{1/2}$ --- to observable galaxy properties -- like galaxy colour or star formation rate \citep{hearin2013darkside,hearin2014dark,watson2015predicting}.
% Traditional AM and how it connects to CAM, one equation explanation of AM and CAM
It is similar to the traditional abundance matching algorithm \citep{kravtsov2004dark}, which assigns stellar masses or luminosities to simulated dark matter haloes. Traditional abundance matching evaluates the function $N_\star^{-1}(N_{\rm dm}(M_{\rm vir}))$, where $N_\star$ and $N_{\rm dm}$ are some observed cumulative stellar mass function and theoretical cumulative mass function, respectively. 
Similarly, CAM assigns galaxy properties via $F_{\rm gal}^{-1}(F_{\rm halo}(\xmah|M_\star)|M_\star)$, where $F_{\rm halo}$ and $F_{\rm gal}$ are the conditional CDFs at a fixed stellar mass $M_\star$ for some theoretical tracer of halo age, $\xmah,$ and the CDF for the target observable galaxy property, respectively.

The primary application of CAM is generating empirical models of observable properties. But more generally, CAM is a method that optimally implements a specific assumption for the connection between halo growth and halo/galaxy properties: a given halo property $\yhalo$ is entirely and monotonically determined by a given feature of a halo's MAH $\xmah$. If this assumption is correct, CAM predictions will be the exact values of the given halo property, and failures in this assumption propagate into inaccuracies in CAM predictions. Therefore, throughout this paper, we use the CAM prediction strength as a measure of how well a given halo property $\yhalo$ can be understood to be determined by a given proxy of halo growth $\xmah$. Moreover, multi-parameter models which have improved predictive power over CAM are evidence that the halo property in question is influenced by multiple features in a halo's MAH.

In this work, the CAM algorithm is used to abundance match a given MAH feature $\xmah$ to a given present-day halo property $\yhalo$, at fixed present-day halo mass $\mvir$. Specifically with the equation: 
\begin{equation}
    \yhalo = F_{\rm halo}^{-1}(F_{\rm mah} (\xmah | \mvir) | \mvir)
\end{equation}
where $F_{\rm halo}$ and $F_{\rm mah}$ are the conditional CDF of the present-day halo property and the MAH feature respectively. Throughout, we condition at a fixed mass bin equal to the one used for constructing the  $\simsample$ data set. 

We pick the MAH property $\xmah$ for abundance matching to be the scale $a(\mopt)$ (see equation \ref{eq:am-ma-relationship}) at a fixed mass bin $\mopt$ that optimally correlates with $\yhalo$ across all $m$. For example, when $\yhalo = \cvir$, we find $\mopt \approx 0.5$ in our \simsample data set, so that $\xmah = a(\mopt) \approx a(0.5) = a_{1/2}$. The optimal mass bin $\mopt$ satisfies the equation:
\begin{equation}
    \max \, (\rhosp(a(m),\yhalo)) = \rhosp(a(\mopt),\yhalo).
\label{eq:mopt}
\end{equation}
Similarly, we could have chosen $\xmah = m(\aopt)$ where the optimal scale $\aopt$ satisfies: 
\begin{equation}
    \max \, (\rhosp(m(a),\yhalo)) = \rhosp(m(\aopt),\yhalo),
\label{eq:aopt}
\end{equation}
but we find that $a(\mopt)$ has overall higher correlations across all halo properties than $m(\aopt)$. We refer to the algorithm that uses $a(\mopt)$ to abundance match between MAH and halo properties at a given halo mass as `\camopt'. We use \camopt to predict a given halo property $\yhalo$ from the MAH of a halo in \cref{sec:pred-present-with-mah}. 

CAM is a simple, yet powerful empirical non-parametric approach to matching any pair of strongly correlated variables. It however has some important limitations: (1) It is unable to match multiple variables to another set of multiple variables. (2) It does not incorporate the scatter between prediction and target when matching. We address these limitations of CAM in the algorithms described next.

%---------------------------------------------------------
\subsubsection{MultiCAM} \label{sec:multicam-general}
%---------------------------------------------------------

We propose the new algorithm \textit{\multicam} to address these limitations of CAM. \multicam generalizes CAM to match multiple MAH properties to multiple present-day halo properties simultaneously. To accomplish this, \multicam first introduces a multi-variable linear regression between the multiple features and target variables. Then, \multicam marginally matches the distribution of outputs to the true distribution of targets.
In our context, different halo properties correlate more or less strongly at different time scales of a halo's growth history \citep[e.g.][]{wong2012dark}. This means that matching multiple variables consistently is essential for exploring the connections in this work. 

\multicam also includes a pre-processing step where all features and target variables are marginally transformed to Gaussian distributions. At the end of the procedure, all variables are transformed to their original space. This pre-processing step is beneficial in the context of linear regression since it allows for a version of \multicam that introduces scatter between the features and targets, as discussed in detail in Sections \ref{sec:lr-and-gaussian-sampling} and \ref{sec:multicam-scatter}. 

The \multicam algorithm is illustrated in \cref{fig:cartoon} and in detail consists of the following: 
\begin{enumerate}
    \item Collect all desired features for prediction, $\Xv$, and targets, $\Yv$, from a given data set. 
    For example, $\Xv$ can be set to the full MAH of all haloes in the data set: $\Xv_{\rm mah} = \{a(m_{i})\}_{i=1}^{N}$, where $\{m_{i}\}_{i=1}^{N}$ are some pre-defined mass bins with $m_{N} = 1$. Similarly, $\Yv$ can be set to all the halo present-day properties we consider in this work $\Yv_{\rm halo} = \{\cvir, \ToverU, \xoff, \spin, c/a\}$, which are described in \cref{sec:halo-properties}. 

    \item We marginally transform each individual feature from its empirical distribution to a normal distribution (top left of figure) to a Gaussian distribution. We do this via the inverse transform method \citep[e.g.][]{devroye1986sample}, which can map any 1D data set of variables to have any other desired empirical distribution without changing the rank-ordering of its points. 

    \item We then take the subset of marginalized Gaussian features $\tilde{\Xv}_{\rm train}$ and targets $\tilde{\Yv}_{\rm train}$ in the training set, and train a linear regression model for prediction in this Gaussianized space for these features and targets.
    
    \item We then use the marginalized Gaussian features in the testing set $\tilde{\Xv}_{\rm test}$ and apply linear regression to obtain the corresponding set of predictions $\tilde{\Yv}_{\rm pred}$. 

    \item The predictions from the linear regression model $\tilde{\Yv}_{\rm pred}$ are not guaranteed to follow the distribution of transformed targets (they tend to be narrower). 
    Thus, we apply one more quantile transformer to $\tilde{\Yv}_{\rm pred}$ and make its distribution (marginally) Gaussian, which then matches the distribution of the Gaussianized training targets $\tilde{\Yv}_{\rm train}$. This is illustrated in the bottom-left corner of \cref{fig:cartoon}. 

    \item Finally, we transform the Gaussianized predictions $\tilde{\Yv}_{\rm pred}$ back into original target space by applying the inverse of the original quantile transformer used to map training target variables to the Gaussianized space. The result is the final \multicam prediction $\Yv_{\rm pred}$. 
\end{enumerate}
This approach incorporates the multi-variable prediction accuracy from linear regression while preserving the properties of the marginal predictor distributions. 
Due to the quantile transformations illustrated in \cref{fig:cartoon}, our procedure automatically outputs predictions whose marginal distributions match the marginal distributions of the training data. This means that the outputs from \multicam have a correlation strength with the true targets that is at least as high as CAM (see \cref{sec:mah-correlations}). In fact, \multicam exactly reduces to CAM in the case of 1D features and targets. In summary, \multicam also has the added advantage of (1) predicting multiple target properties from multiple input properties and (2) taking advantage of the increased accuracy from linear regression. 

This version of \multicam that uses linear regression still faces one key limitation in that it doesn't account for the scatter between features and targets, and thus will not reproduce the correct correlations between output properties. To address this, we first discuss the relationship between linear regression and sampling from a conditional Gaussian. Second, we discuss a method that maintains the correlation between sampled properties that is based on using conditional Gaussian sampling within MultiCAM instead of linear regression. 

%---------------------------------------------------------
\subsubsection{Linear regression and conditional Gaussian sampling} \label{sec:lr-and-gaussian-sampling}
%---------------------------------------------------------
We start by discussing the theoretical framework of conditional Gaussian prediction, and then connect it with linear regression and \multicam. Assume that you have some multi-dimensional features $X$ and multi-dimensional targets $Y$ that are jointly distributed as a multivariate Gaussian $P_{X,Y}$. Given a new feature test point $x^{\star}$, we consider the conditional distribution $P_{Y|x^{\star}}$ in order to choose our new prediction based on $x^{\star}$. The conditional distribution $P_{Y|x^{\star}}$ is also Gaussian with mean $\bar{\mu}(x^{\star})$ and covariance matrix $\bar{\Sigma}$. The equations to derive the conditional parameters $\bar{\mu}(x^{\star})$ and $\bar{\Sigma}$ from empirical estimates of the joint distribution $P_{X,Y}$ parameters can be found in \Cref{app:multivariate-gaussian-algorithm}.

Given this framework, there are two different goals we could choose to pursue: (1) Minimize (squared) residuals of the prediction $Y_{\rm pred}(X)$ relative to the target $Y$ or (2) Sample points $Y$ such that their distribution matches the true target distribution $P(Y)$, including in its correlations between different target variables.

The first goal is achieved by using the mode of the conditional distribution directly as the prediction: 
\begin{equation}
    y_{\rm pred}(x^{\star}) \equiv \bar{\mu}(x^{\star}).
\label{eq:mode-of-gauss-conditional}
\end{equation}
Based on the expression for $\bar{\mu}(x^{\star})$ in \cref{eq:mu_bar}, we can see how this prediction would not take into account the intrinsic scatter of the target distribution, as there is no term with $\Sigma_{yy}$ -- the covariance matrix between target variables. 

The second goal can be achieved by sampling the conditional distribution $P_{Y \vert X}$ after sampling $P(X)$. Concretely, given a test point $x^{\star} \sim P(X)$, we choose as our prediction samples directly from the conditional normal distribution $P_{Y | x^{\star}}$: 
\begin{equation}
    y_{\rm pred}(x^{\star}) \sim \normal{\bar{\mu}(x^{\star})}{\bar{\Sigma}}.
\label{eq:gauss-multi-sampling}
\end{equation}
This second approach does incorporate the intrinsic scatter in the target distribution as $\bar{\Sigma}$ depends on $\Sigma_{yy}$, as can be seen in \cref{eq:bar-sigma} in \cref{app:multivariate-gaussian-algorithm}. We denote this approach \textit{conditional Gaussian sampling}.

In \cref{app:lr-is-special}, we prove that the mode of the conditional distribution $P_{Y | X}$ (equation \ref{eq:mode-of-gauss-conditional}) is equivalent to the linear regression output if $X,Y$ are jointly normal distributed. Additionally, \multicam already includes a pre-processing step (step 2 of the algorithm in \cref{sec:multicam-general}) where we try to bring features $X$ and targets $Y$ close to a joint Gaussian. These two facts combined imply that the conditional Gaussian sampling approach (equation \ref{eq:gauss-multi-sampling}) is a natural replacement for the linear regression prediction algorithm within \multicam that could allow us to account for the scatter between targets. 

Finally, note that we restrict analysis in this paper to simulation data, where we can train the entirety of $\Sigma$ and account for the explicit covariance between all features and predicted quantities. 
However, conditional Gaussian sampling provides an avenue to use \multicam as an interpretable empirical model. In the simplest case, if we consider traditional CAM as such an empirical model, the ``fit'' procedure would consist of $\Sigma$ containing one row for $X_{\rm mah}$, one row for the target galaxy observable, and off-diagonal terms artificially fixed to assume perfect correlation. 
In the more general case using \multicam with conditional Gaussian sampling, we would perform an analogous ``fit'' procedure by taking any subset of the elements in $\Sigma$ as free parameters.

\begin{table}
\centering
\begin{tabular}{|c||c|c|c|}
\hline
Model & $x_{\rm off}$,\,$\lambda_{\rm bullock}$ & $x_{\rm off}$,\,$c/a$ & $\lambda_{\rm bullock}$,\,$c/a$ \\ [0.5ex]
\hline\hline
\rm True & 0.51 & -0.43 & -0.29 \\
\hline
\rm CAM $a(m_{\rm opt})$ & 0.62 & -0.79 & -0.88 \\
\hline
\rm MultiCAM (no scatter) & 0.93 & -0.96 & -0.95 \\
\hline
\rm MultiCAM (with scatter) & 0.50 & -0.45 & -0.31 \\
\hline
\end{tabular}
\caption{
    \textbf{Correlations between halo properties predicted from each model}. We show the Spearman correlation between each pair of predicted target $z=0$ halo properties given their MAH using three different methods. The training and testing (`True') data sets are equivalent to the one used for \cref{fig:3panel-triangle} in \cref{sec:multicam-scatter}.
}
\label{tab:3panel-correlation}
\end{table}

%---------------------------------------------------------
\subsubsection{MultiCAM with scatter} \label{sec:multicam-scatter}
%---------------------------------------------------------
As mentioned previously, the \multicam algorithm presented in \cref{sec:multicam-general} cannot correctly capture the correlation between targets. As explained in \cref{sec:lr-and-gaussian-sampling} this is because the prediction model connecting features and targets, linear regression, does not account for the scatter in the target distribution. 

However, given that the \multicam algorithm presented in \cref{sec:multicam-general} already includes a normalizing pre-processing step (step 2), we can replace the prediction model from linear regression (step 3 and 4) to conditional Gaussian sampling (equation \ref{eq:gauss-multi-sampling}) to solve this problem. As explained in \cref{sec:lr-and-gaussian-sampling}, the pre-processing step allows us to interpret this replacement as using the same joint normal distribution to solve a different goal, that of directly sampling $P(Y)$. This can be achieved by using the conditional Gaussian sampling approach within \multicam, since we will be explicitly incorporating the scatter between targets in our predictions. Therefore, for the rest of this subsection, we denote this new version of \multicam as \textit{\multicam (with scatter)} to distinguish it from the method in \cref{sec:multicam-general} which we will denote as \textit{\multicam (no scatter)}. Unless otherwise stated, in the rest of the paper `\multicam' refers to `\multicam (no scatter)'. 

Importantly, the \multicam (with scatter) approach explicitly models scatter between features and targets, i.e. a given test data point of features can be used to sample multiple predictions from the conditional normal distribution. This means that the point estimate accuracy of \multicam (with scatter) will be lower compared to \multicam (no scatter), since we are introducing noise into the prediction. However, we will show how this simple extension allows for capturing the lion's share of the covariance between variables while still matching the marginal distributions exactly.

\begin{figure*}
    \centering
    \begin{subfigure}
        \centering
        \includegraphics[width=0.65\columnwidth, trim = 0 0 130 80, clip]{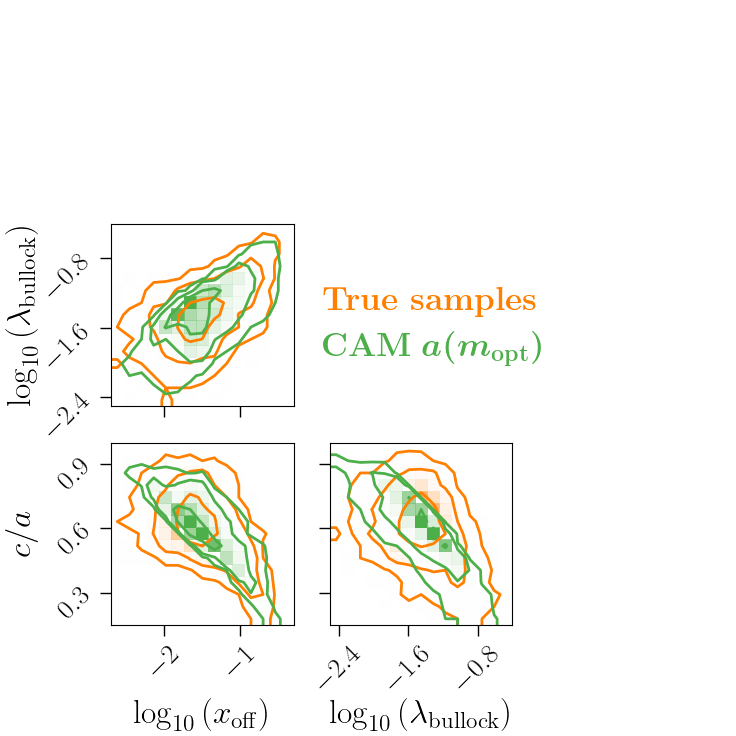}
    \end{subfigure}
    \quad
    \begin{subfigure}
        \centering
        \includegraphics[width=0.65\columnwidth, trim = 0 0 130 80, clip]{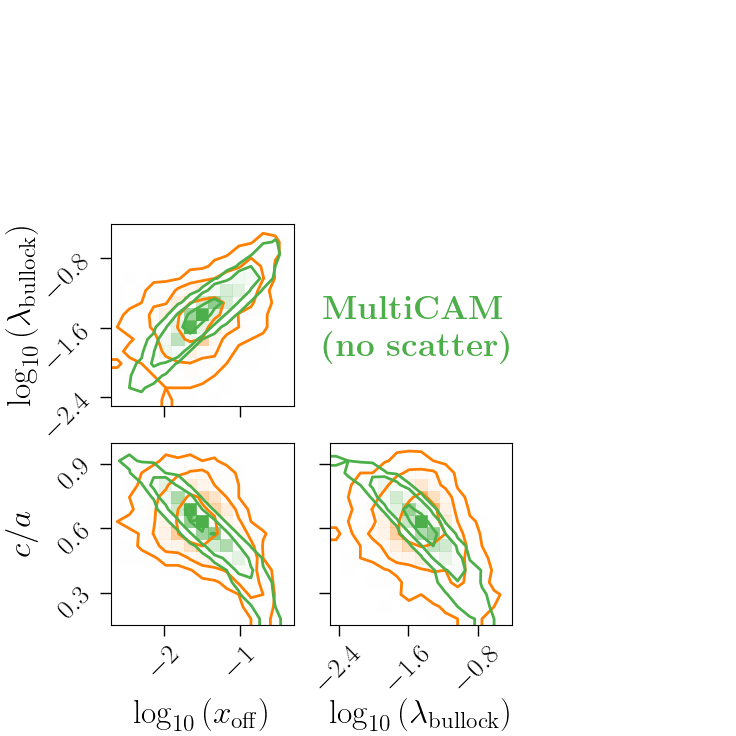}
    \end{subfigure}
    \quad
    \begin{subfigure}
        \centering
        \includegraphics[width=0.65\columnwidth, trim = 0 0 130 80, clip]{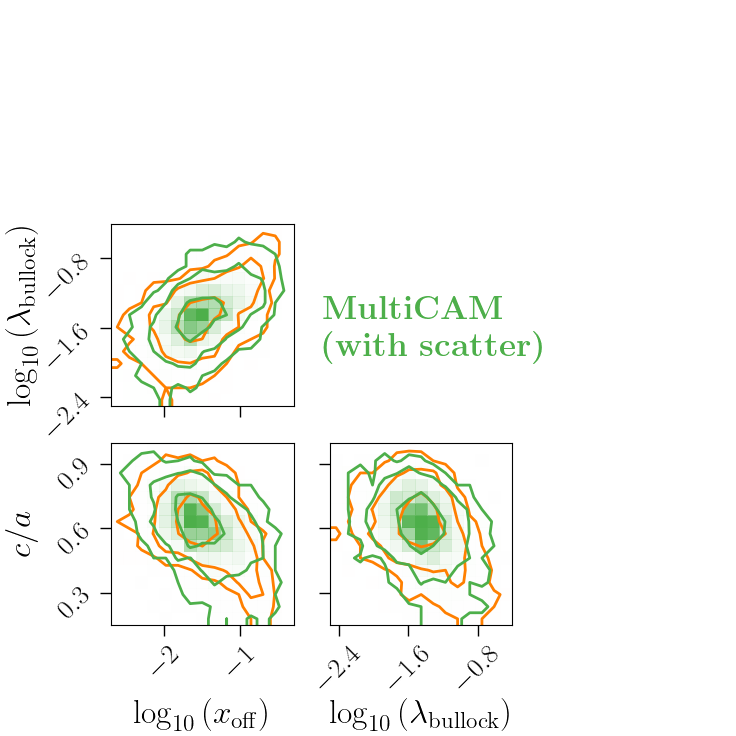}
    \end{subfigure}
    \caption{
        \textbf{2D Scatter with contours from samples of $z=0$ halo properties comparing different models.} We show plots of $1\sigma$, $2\sigma$, and $3\sigma$ contours for the 2D histograms of $3000$ samples of $z=0$ halo properties given their MAH using three different methods. Each method is applied to $\spin, c/a,$ and $\xoff$ within our \simsample data set.
        The orange contours in each subplot show the true empirical distributions of these properties.
        The green contours of each subplot were produced by applying three different prediction methods to these halo properties: \camopt (left), \multicam with no scatter (middle), and \multicam with scatter (right). These models were trained on the remainder of the \simsample data set. For more details on the different methods used, see \cref{sec:statistical-algorithms}, and for more discussion on the figure see \cref{sec:multicam-scatter}.
    }
    \label{fig:3panel-triangle}
\end{figure*}

To demonstrate this, we first train each of the models presented so far --- \camopt, \multicam (no scatter), and \multicam (with scatter) --- on 7000 random haloes from the $\simsample$ data set using the full MAH $\{a(m_{i})\}_{i=1}^N$ of each halo as features and three present-day halo properties as targets: $c/a, \spin$, and $\xoff$.
The three models are then tested on full MAH of remaining 3000 haloes from the $\simsample$ data set and the 2D, $1\sigma$, $2\sigma$, and $3\sigma$ contours between each pair of target predicted variables are plotted as shown in \cref{fig:3panel-triangle}. 
The true contours are shown in orange and the predicted contours by each model in green. 

In \cref{fig:3panel-triangle}, we see that \camopt and \multicam (no scatter) fail to match the 2D distributions of halo properties. 
For \camopt, the width of the green contours in each panel directly corresponds to the covariance between the $a(\mopt)$ of each property, since CAM does a one-to-one matching between these. 
For example, $\xoff$ and $\spin$ are the target variables with the largest difference in their corresponding $\mopt$, as shown in \cref{tab:optimal-corrs}. \cref{fig:correlation-matrix-am} demonstrates that a larger difference in mass bins $m$ between a pair of scales $a(m)$ implies a lower covariance between them. Thus, we expect a weaker correlation between the \camopt-predicted $\xoff$ and $\spin$ than for the other pairs of variables. This is exactly what we see in the leftmost subplot in \cref{fig:3panel-triangle}.
\multicam (no scatter) has the narrowest contours out of the three methods. This is because the predicted variables use the same sets of MAHs and there is substantial overlap in  the relative importance of different epochs (see \cref{sec:mah-correlations}).
However, \multicam (with scatter) has contours that seem to match the true contours more closely. 

Additionally, \cref{tab:3panel-correlation} shows the correlation between each pair of $z=0$ halo properties for each of the three models. We can quantitatively reach the same conclusions suggested by \cref{fig:3panel-triangle}: the correlations between target properties outputted by \multicam (with scatter) agree closely with the true correlations, but this is not the case for CAM $a(m_{\rm opt})$ and \multicam (no scatter).

The full triangle plot applying \multicam (with scatter) to the main present-day properties considered in this work is shown in \cref{fig:full-triangle} of \cref{app:triangle-covariances}, which shows good agreement in both 1D marginals and 2D contours. As explained in \cref{sec:lr-and-gaussian-sampling}, \multicam (with scatter) can successfully capture the covariance between target variables since the sampling scatter depends directly on this covariance (equation \ref{eq:gauss-multi-sampling}). 

In summary, \cref{tab:3panel-correlation}, \cref{fig:3panel-triangle}, and \cref{fig:full-triangle} demonstrate that \multicam (with scatter) can be used to successfully emulate present-day halo properties given the full MAH of a dark matter halo.

%%%%%%%%%%%%%%%%%%%%%%%%%%%%%%%%%%%%
\section{Results}\label{sec:results}
%%%%%%%%%%%%%%%%%%%%%%%%%%%%%%%%%%%%

In this section, we focus on understanding the statistical properties of our \simsample data set through correlations and evaluate the \multicam approach. We choose to focus on the following $z=0$ halo properties for our analysis: concentration, $\cvir$, virial ratio, $\ToverU$, centre of mass displacement, $\xoff$, spin Bullock, $\spin$, and second minor-axis to major-axis ratio, $c/a$. 

We analyse the \simsample data set as defined in \cref{sec:datasets}. We divide the \simsample halo sample into a training set of 7000 haloes and a test set of 3000 haloes (unless otherwise stated). The performance metrics of trained models are evaluated only on the test set. The error bars reported in all our results are standard errors estimated from jackknife resampling over $8$ equal volume subcubes of the simulation.

%======================================================
\subsection{Autocorrelation of halo MAHs}\label{sec:autocorrmah}
%======================================================'

\begin{figure}
    \centering
    \begin{subfigure}
        \centering
        \includegraphics[width=\columnwidth]{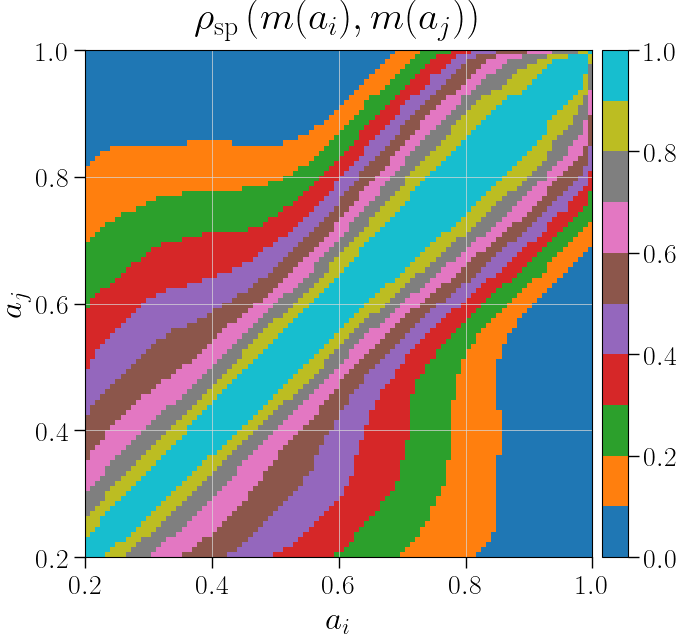}
    \end{subfigure} 
    \par \vspace{-0.25\baselineskip}
    \begin{subfigure}
        \centering
        \includegraphics[width=\columnwidth]{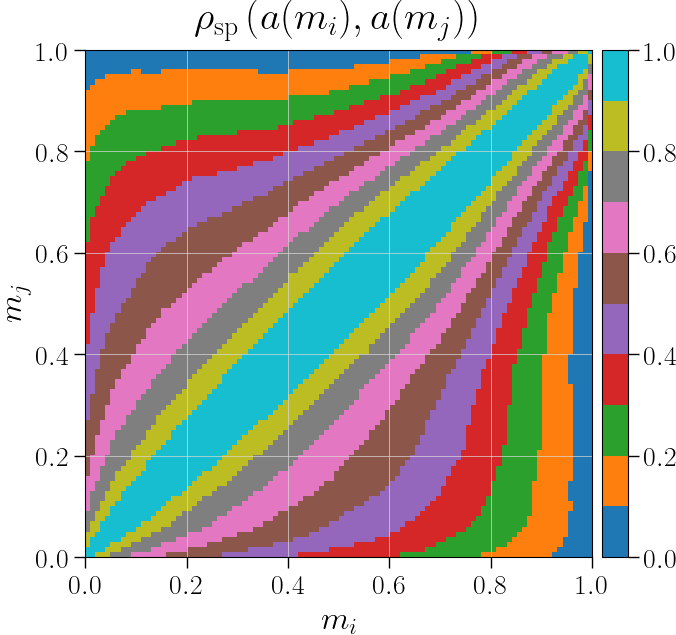}
    \end{subfigure}
    \vspace{-0.5cm}
    \caption{
        \textbf{Internal MAH Spearman correlation.} The colour in each 2D bin (pixel) of these plots corresponds to the Spearman correlation $\rhosp(m(a_{i}), m(a_{j}))$ between the mass fraction at a given pair of formation times, $(a_{i}, a_{j})$ (top), and the Spearman correlation $\rhosp(a(m_{i}), a(m_{j}))$ between the formation time at a given pair of mass fractions, $(m_{i}, m_{j})$ (bottom), for all the $10000$ haloes in our \simsample data set. See \cref{sec:mah-correlations} for additional discussion.
    }
    \label{fig:correlation-matrix-am}
\end{figure}

\cref{fig:correlation-matrix-am} shows 2D histograms where we colour code each pixel (bin) by Spearman correlation strength. The top plot shows the Spearman correlation, $\rhosp(m(a_{i}),\,m(a_{j}))$, between mass fraction at a given pair of formation times. The bottom plot shows the Spearman correlation, $\rhosp(a(m_{i}),\,a(m_{j}))$, between the formation time at a given pair of mass fractions in our \simsample data set. 

In the top panel, we see that $m(a)$ values are strongly correlated with one another for small ($\Delta a\approx 0.1$) changes in $a.$ Similarly, in the bottom plot we see that $a(m)$ values are strongly correlated with one another for small  ($\Delta m \approx 0.1$) changes in $m$. This suggests that we can achieve a similar prediction accuracy with a sparser subset of the MAH information. For example, if we wanted to retain information at a level of $\rhosp \sim 0.9$ between adjacent bins, we could choose data at approximately a spacing of $\Delta a = 0.05$ which would result in approximately ten times less data. 

Another takeaway from the top plot is that adjacent snapshots at both early and late times are strongly correlated (see \cref{sec:sim-suite}). The distinct output cadence of Bolshoi should therefore have minimal impact in the following analysis.

The takeaways for the bottom plot are similar to those from the top plot.

%======================================================
\subsection{Correlations of MAH and present-day halo properties} \label{sec:mah-correlations}
%======================================================'

\begin{figure*}
    \centering
    \begin{subfigure}
        \centering
        \includegraphics[width=\columnwidth]{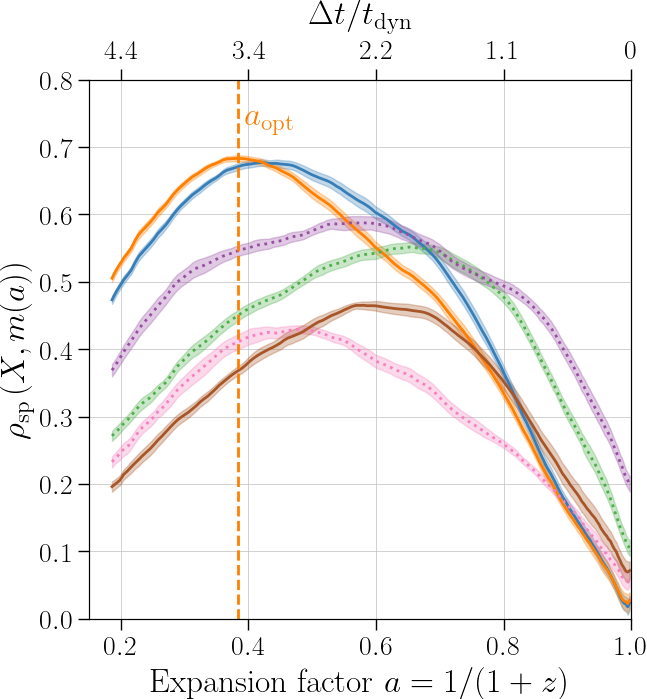}
    \end{subfigure} 
    \quad
    \begin{subfigure}
        \centering
        \includegraphics[width=\columnwidth]{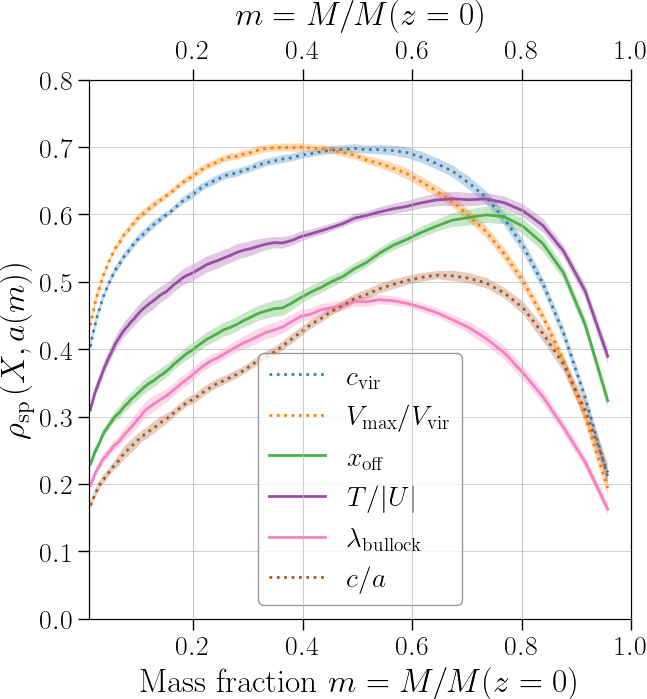}
    \end{subfigure}
    \caption{
        \textbf{Correlation of accretion history with present-day properties.} We show the Spearman correlation coefficient between different present-day halo properties, $X$, and accretion history, parametrized as $m(a)$ or $a(m)$. The correlation is calculated based on our complete $10000$ halo sample \simsample.
        The coloured bands around each curve show the error estimated by jackknife resampling. In both figures, solid lines indicate positive correlation value and dotted lines indicate negative correlation value.
        See \cref{sec:mah-correlations} for additional discussion on these plots. See \cref{tab:optimal-corrs} for the specific values of optimal correlations between halo properties and MAH (peaks in these plots). 
        The annotated orange dashed vertical line in the left-hand plot illustrates one such optimal correlation $\aopt$ for the $\vmaxovervvir$ property (whose exact value is the second row in \cref{tab:optimal-corrs}).
    } 
    \label{fig:mah-correlations-with-properties}
\end{figure*}

In \cref{fig:mah-correlations-with-properties}, we show the Spearman correlation coefficient between several present-day halo properties and the halo accretion history, parametrized as $m(a)$ (left) and $a(m)$ (right).  We compute the correlation using all $10,000$ haloes in our \simsample sample. The coloured red bands correspond to the uncertainty on each curve as estimated by jackknife resampling. In this figure, solid lines are used to represent positive correlation values and dotted lines represent negative values.

Both figures illustrate that present-day halo properties contain information about the growth of haloes back to very early times, $z\approx 4,$ and at times when haloes were $\approx$10\% to 20\% of their current mass. As expected, formation times correlate positively with $c_{\rm vir}$ \citep{wechsler2002concentrationsfromassembly}, $V_{\rm max}/V_{\rm vir}$ (this follows directly from the $c_{\rm vir}$ correlation with growth), $c/a$ \citep{allgood2006shape,chen19}, and negatively with $T/|U|,$ $x_{\rm off}$ \citep{maccio2007concentration}, and $\lambda_{\rm bullock}$ \citep{vitvitska2002spin}.

Inner halo structure, tracked by $c_{\rm vir}$ and $V_{\rm max}/V_{\rm vir}$, most strongly correlates with early times, $\approx 3.4\tau_{\rm dyn}$ in the past, when haloes were roughly half their current mass. This is consistent with models of halo structure in which the inner profile is  primarily set by long term growth trends  \citep[e.g.][]{dalal2010origin,ludlow2013massprofileL}. More recently, \citet{wang2020concentrations} systematically examined the correlation between the present-day concentration and different stages of halo mass assembly.
They found that there are extended periods in the assembly history that correlate strongly with the present-day halo structure, which justifies the use of various definitions of halo formation time with which to predict present-day concentrations.
These findings are qualitatively consistent with our results. 

The other properties that we track, $\xoff$, $\ToverU$, $\lambda_{\rm bullock}$, and $c/a$ have relatively larger predictive power at late times compared with properties that more closely describe the halo inner structure, such as $\cvir$. All four are expected to be tracers of dynamically unrelaxed haloes that have recently experienced major mergers or rapid, anisotropic smooth accretion from nearby filaments. More relaxed haloes will be more spherical, more centred on its most bound point, and will have a virial ratio closer to 0.5 \citep{mo2010thebook}. 
Any deviations would be caused by recent external influences, which are typically mergers for non-subhaloes (although mass loss due to tidal stripping can also influence halo properties, e.g., \citealp{tucci2021spinbias}). 
The correlation with spin is generally understood to arise because a slowly accreting halo will generally accrete isotropically, reducing its normalized angular momentum over time, while a rapidly accreting halo will experience larger mergers which will inject large amounts of angular momentum into the system \citep[e.g.][]{vitvitska2002spin}. However, halo spin also plays a large role in the early collapse of dark matter perturbations prior to forming haloes \citep[e.g.][]{sheth2001ellipsoidal}, meaning that it should not be thought of as a purely late-time phenomenon. 

\cref{tab:optimal-corrs} contains the values of optimal correlations between halo properties and MAH which correspond to the peaks of the curves in \cref{fig:mah-correlations-with-properties}. As an example, we include a dashed vertical line in the left-hand panel of \cref{fig:mah-correlations-with-properties} which intersects the peak of the correlation curve for the $\vmaxovervvir$ property. In other words, the $x$-value of the vertical orange line is $\aopt$ when $X = \vmaxovervvir$, which corresponds to the second row of \cref{tab:optimal-corrs}.

We also measured correlations with other measures of triaxiality, $q$ and semi-minor axis ratio $b/a$. The $\aopt$ of $q$ and the ellipticity ratio $c/a$ are the same, but $q$ has a slightly higher peak absolute Spearman correlation with MAH of $|\rhosp| = 0.533$ compared to $|\rhosp| = 0.510$ for $c/a$. The correlation between $b/a$ and MAH is comparable with that between $c/a$ and MAH.

Analogously, we evaluated the results for $\lambda_{\mathrm{Peebles}}$, the Peebles spin parameter.  This measurement was comparable with $\lambda_{\mathrm{Bullock}}$, but $\lambda_{\mathrm{Bullock}}$ has a higher peak correlation of $\rhosp=0.473$ compared to $\lambda_{\mathrm{Peebles}}$, which has a peak correlation of $\rhosp = 0.384$, likely due to the fact that measurements of internal energy for $\lambda_{\mathrm{Peebles}}$ are less stable leading to weaker signals.  We use $\lambda_{\mathrm{Bullock}}$ in all subsequent analyses considering the spin of the haloes.

In comparing $V_{\rm max}/V_{\rm vir}$ to $\cvir$, the peak correlation occurs slightly earlier in the former quantity with comparable correlation strength.  

Finally, we note that the correlation between $m(a)$ and present-day properties drops for both very early times and late times (left side of \cref{fig:mah-correlations-with-properties}). The former drop is because halo properties across the board are no longer correlated with those early times. The latter drop is related to the distribution of $m(a)$ at later times. The limiting behaviour of all $m(a)$ curves is to approach $1$ as $a$ approaches $1$. This means that the limiting behaviour of the scatter in $m(a)$ is to approach zero as $a$ approaches 1 and, conversely, to grow as $a$ decreases. This trend of decreasing scatter in $m(a)$ with increasing $a$ is explained by the fact that once a halo reaches or exceeds its present-day mass, $m(a)$ becomes fixed to $1$ for the rest of that halo's history (see equation \ref{eq:mass-fraction}). However, if we think of $m(a)$ as an accretion rate between the $a$ and the present-day mass, as the time range over which the accretion rate is measured decreases, the intrinsic noise in that measurement becomes larger. The width of $m(a)$ eventually becomes smaller than this intrinsic noise in $m(a)$, thus decreasing the correlation. 
The cause of the drop in correlation between $a(m)$ and the halo properties (right side of \cref{fig:mah-correlations-with-properties}) at very low and very high mass fractions stems from a similar argument.

\begin{table*}
\centering
\begin{tabular}{|c|c|c|c|c|}
\hline
$X$ & $a_{\rm opt}$ & $\rho_{\rm sp}\left(X, m(a_{\rm opt})\right)$ & $m_{\rm opt}$ & $\rho_{\rm sp}\left(X, a(m_{\rm opt})\right)$ \\ [0.5ex]
\hline\hline
$c_{\rm vir}$ & $0.420 \pm 0.006$ & $0.677 \pm 0.005$ & $0.495 \pm 0.021$ & $-0.698 \pm 0.006$\\ \hline
$V_{\rm max} / V_{\rm vir}$ & $0.384 \pm 0.018$ & $0.683 \pm 0.004$ & $0.397 \pm 0.051$ & $-0.700 \pm 0.006$\\ \hline
$V_{\rm off} / V_{\rm vir}$ & $0.616 \pm 0.040$ & $-0.523 \pm 0.008$ & $0.735 \pm 0.036$ & $0.574 \pm 0.008$\\ \hline
$x_{\rm off}$ & $0.652 \pm 0.007$ & $-0.551 \pm 0.007$ & $0.735 \pm 0.032$ & $0.599 \pm 0.012$\\ \hline
$T/|U|$ & $0.562 \pm 0.052$ & $-0.588 \pm 0.010$ & $0.673 \pm 0.082$ & $0.623 \pm 0.011$\\ \hline
$\lambda_{\rm peebles}$ & $0.480 \pm 0.010$ & $-0.342 \pm 0.007$ & $0.541 \pm 0.023$ & $0.384 \pm 0.007$\\ \hline
$\lambda_{\rm bullock}$ & $0.480 \pm 0.007$ & $-0.430 \pm 0.006$ & $0.541 \pm 0.023$ & $0.473 \pm 0.007$\\ \hline
$c/a$ & $0.580 \pm 0.025$ & $0.465 \pm 0.006$ & $0.644 \pm 0.033$ & $-0.510 \pm 0.006$\\ \hline
$b/a$ & $0.592 \pm 0.006$ & $0.460 \pm 0.008$ & $0.673 \pm 0.029$ & $-0.505 \pm 0.008$\\ \hline
$q$ & $0.580 \pm 0.022$ & $0.487 \pm 0.006$ & $0.673 \pm 0.033$ & $-0.533 \pm 0.007$\\ \hline
$R_{200m} / R_{\rm vir}$ & $0.544 \pm 0.008$ & $-0.367 \pm 0.007$ & $0.292 \pm 0.045$ & $0.321 \pm 0.008$\\ \hline
$R_{500c} / R_{\rm vir}$ & $0.532 \pm 0.011$ & $0.593 \pm 0.011$ & $0.673 \pm 0.029$ & $-0.614 \pm 0.011$\\ \hline
\end{tabular}
\caption{
    \textbf{Optimal correlations 
    between present-day halo 
    properties and single-epoch
    measurements of the MAH.}
   In this table we show the optimal scale factors, $a_{\rm opt},$ and mass fractions, $m_{\rm opt}$, at which the present-day halo properties of our halo sample achieve their maximum absolute Spearman correlation with $m(a)$ or $a(m)$ respectively.
   These values correspond to the maxima of the curves in \cref{fig:mah-correlations-with-properties}. The precise definition of $\mopt$ and $\aopt$ can be found in \cref{sec:mah-summaries}. 
}
\label{tab:optimal-corrs}
\end{table*}

%======================================================
\subsection{Predictions of present-day properties based on MAH} \label{sec:pred-present-with-mah}
%======================================================

\begin{figure}
    \centering
    \includegraphics[width=\columnwidth]{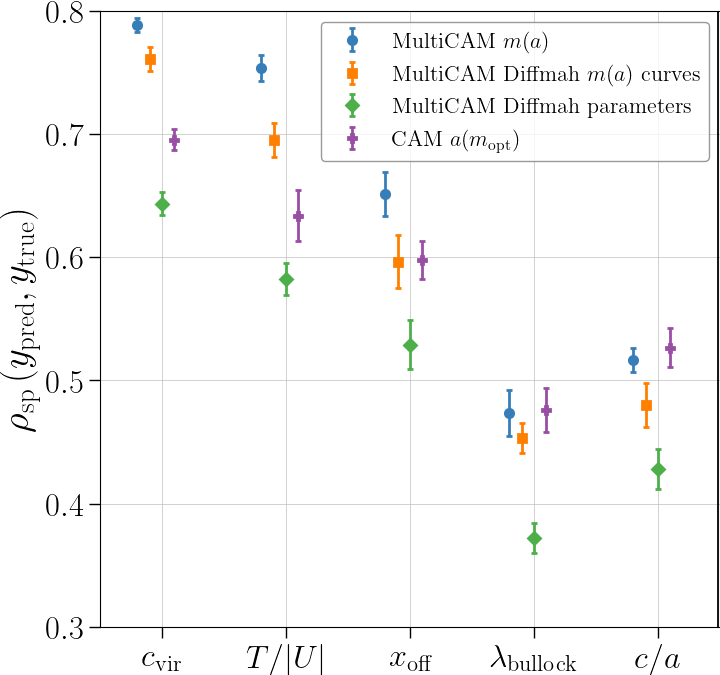}
    \caption{
        \textbf{Correlation between predictions of $z=0$ halo properties based on MAH and true properties.} We show the Spearman correlation between several true $z=0$ halo properties and predicted $z=0$ halo properties using four trained models on the \simsample data set. The first three models are based on MultiCAM trained on full MAH (blue circle), on \diffmah curve fits to the MAH curves evaluated at the same scale factors as the full MAH (orange square), and on the parameters of the \diffmah fit (green diamond). The last model (purple plus) is the prediction of the CAM algorithm using the corresponding $a(m_{\rm opt})$ (defined in equation \ref{eq:mopt}) for each halo property. See \cref{sec:pred-present-with-mah} for additional discussion on this figure.
    } 
    \label{fig:forward-pred-metrics}
\end{figure}

In \cref{fig:forward-pred-metrics}, we show the Spearman correlation between several predicted halo properties and their true value for four different models described in \cref{sec:statistical-algorithms}.

In blue circles, we show results for our canonical MultiCAM model, using the full MAH of each halo. Under this metric, MultiCAM either outperforms or performs comparably well to the other tested models. 

With orange squares, we show results of applying MultiCAM to the best-fitting \diffmah curve for each MAH (see \cref{sec:diffmah-definitions} for more information). This model is next in predictive power for the target halo properties shown. We highlight the similar performance between this model and MultiCAM trained on the full non-parametrized MAH (blue circles) for most halo properties. The consistency of performance implies that our method leans heavily on information contained within the smooth accretion history.

Next, we show the performance of a model that applies MultiCAM to the three best-fitting parameters from \diffmah in green diamonds.  We note that the \diffmah parameters alone have systematically lower prediction power than the full MAH curve that the \diffmah parameters describe. This may be due to the non-linear mapping of \diffmah parameters onto the MAHs that cannot be captured by the linear modelling we employ in MultiCAM. Further investigation might include testing non-linear models to map \diffmah parameters to halo properties.

Relatedly, the decrease in prediction power for \multicam on \diffmah parameters suggests a degeneracy between \diffmah parameters and present-day halo properties. In that case, the exact parametrization of the \diffmah curve matters. Indeed, one can show from equations \ref{eq:diffmah-base} and \ref{eq:diffmah-exponent} that we can pick a parametrization where we replace $\alpha_{\rm early}$ with $a_{1/2}$ and still get a complete set of \diffmah parameters that uniquely characterizes a MAH curve.
We find an increase of $\geq 0.05$ in the correlation with $\cvir, \spin,$ and $c/a$ with this alternative parametrization. This indicates that the \diffmah parametrization chosen impacts the predictive power of \diffmah parameters, which is also further evidence of the aforementioned degeneracy. 

Finally, in the purple pluses, we show model predictions for CAM evaluated at $a_{\mathrm{opt}}$, which only uses the scale at a single mass fraction of a halo that best correlates with that halo property (see \cref{tab:optimal-corrs}). 
We see that \multicam on the full MAH significantly outperforms \camopt for prediction of most halo properties including: $\cvir, \ToverU,$ and $\xoff$. For the other two halo properties, $\lambda_{\mathrm{bullock}}$ and $c/a$, \multicam and \camopt have (statistically) the same performance. Moreover, CAM performs significantly better than \multicam on \diffmah parameters, which might be related to the fact that CAM $a_{\mathrm{opt}}$ is using (by construction) the best single feature in predicting MAH.

Comparing the individual models within different types of halo properties, we notice a few trends.  First, the full curve from the \diffmah fit performs at least as well as the model trained with CAM $a_{\mathrm{opt}}$ on all halo properties. The MultiCAM on \diffmah fit information provides better predictions on properties that are most strongly correlated with overall MAH, e.g. $\cvir$ and $\ToverU$.  For properties whose predicted values are more weakly correlated with truth (i.e. $\lambda_{\mathrm{Bullock}}$ and $c/a$), all models, except for the one using the \diffmah parameters only, perform similarly. 

%  MultiCAM and aopt
The halo property predictions where CAM applied to $a_{\mathrm{opt}}$ perform comparably well tend to be in the ``worst'' cases of target predictions (e.g. $\xoff$, $\lambda_{\mathrm{Bullock}}$, and $c/a$). We surmise that the comparable performance is due to the fact that these halo properties are largely dependent on the most recent MAH of the haloes and that these properties are even more sensitive to the non-smooth component of the MAH, comprised of moderate and major mergers, which our model does not yet account for.

We additionally investigated whether using the gradient of the MAH could successfully capture the missing major merger information. Specifically, we computed the first-order derivative of MAHs using a Savitzky--Golay Filter \citep{savitzky1964smoothing} and used these derivatives as additional features for \multicam. However, we found no significant difference between \multicam trained on the full MAH and its gradients compared to our canonical \multicam model trained only on the full MAH.

\vspace{-6.5pt}

%======================================================
\subsection{Predictions of MAH summaries based on present-day properties} \label{sec:pred-mah-with-present}
%======================================================

\begin{figure}
    \centering
    \includegraphics[width=\columnwidth]{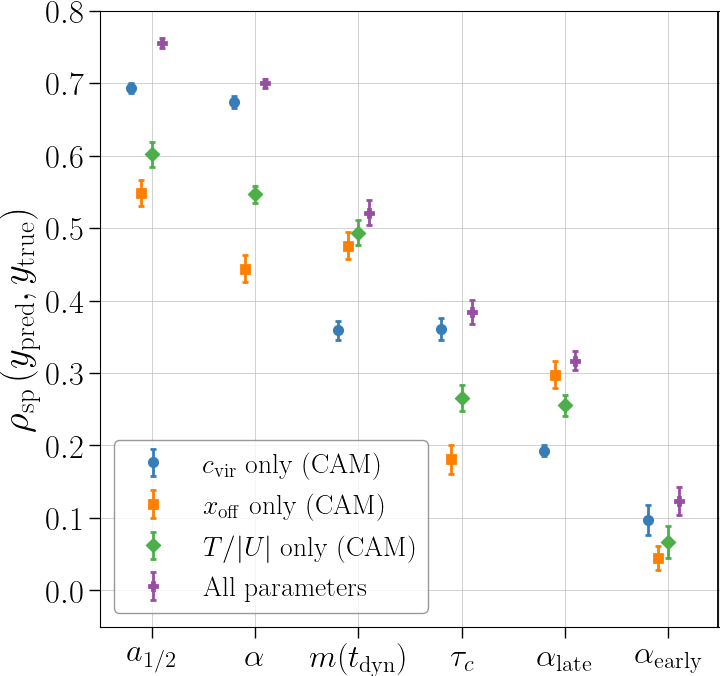}
    \caption{
        \textbf{Correlation between MultiCAM predictions of MAH summaries from $z=0$ halo properties.} Here we show the Spearman correlation between parameters characterizing the MAH of haloes in our testing set, and their predictions using the MultiCAM algorithm trained on subsets of the $z=0$ halo properties. The definitions of these MAH properties can be found in Sections \ref{sec:mah-summaries} and \ref{sec:diffmah-definitions}. 
        The last model (purple cross) corresponds to MultiCAM trained on the following $z=0$ halo properties: $\cvir$, $\vmaxovervvir$, $\xoff$, $\ToverU$, $\spin$, and $c/a$. The correlation from the first three models (blue circle, orange square, and green diamond) is equivalent to the CAM predicted correlation. See \cref{sec:pred-mah-with-present} for additional discussion on this figure.
    }
    \label{fig:inverse-pred-metrics}
\end{figure}

In \cref{fig:inverse-pred-metrics} we use \multicam to perform the inverse of the test shown in \cref{fig:forward-pred-metrics}: predicting summary statistics of a halo's MAH from its $z=0$ halo properties. We attempt to predict $a_{1/2},$ the half-mass scale (equation \ref{eq:a_1/n}), $\alpha$, the characteristic time in an exponential MAH fit (equation \ref{eq:exponental_M(z)}), $m(t_{\rm dyn})$, the accretion rate over a dynamical time (equation \ref{eq:mass-fraction}), and the three \diffmah parameters, $\tau_c,$ $\alpha_{\rm late},$ and $\alpha_{\rm early}$ (equations \ref{eq:diffmah-base} and \ref{eq:diffmah-exponent}). We use \multicam to predict these values with different combinations of $\cvir, \ToverU, \xoff, \spin,$ and $c/a$.

Using \multicam on the full suite of halo properties (purple plus signs) results in strictly more accurate predictions than using a single halo property, as expected. 
As expected from \cref{fig:mah-correlations-with-properties}, $\cvir$ (blue circles) does a better job predicting tracers of early accretion history like $a_{1/2},$ $\alpha,$ and $\alpha_{\rm early}$ than $\xoff$ (orange squares) and $\ToverU$ (green diamonds). The opposite is true for tracers of late accretion history, like $m(\tdyn)$ and $\alpha_{\rm late}$.

Simpler parametrizations, like $\alpha$ for an exponential growth history, are well-predicted, but for more complicated non-linear parameters like in the \diffmah fits, predictions are quite poor. This is most likely due to the degeneracy between \diffmah parameters and halo properties as discussed in \cref{sec:pred-present-with-mah}.

Despite being comparatively poorly predicted, the same trends can be seen in the \diffmah parameters that are seen in the single-epoch MAH tracers. $\alpha_{\rm late}$ encodes behaviour at late times and is better predicted by $x_{\rm off}$ and $T/|U|$ than by $c_{\rm vir}$. $\tau_c$ is sensitive to earlier times and better predicted by $c_{\rm vir}$ than $x_{\rm off}$ and $T/|U|.$ $\alpha_{\rm early}$ probes even earlier times and shows no statistically significant trends. This may either be due to $\alpha_{\rm early}$ having a relatively small impact on the overall MAH or it corresponding to such an early time period that no present-day properties do a good job at tracing it.

\begin{figure*}
    \centering
    \begin{subfigure}
        \centering
        \includegraphics[width=\columnwidth]{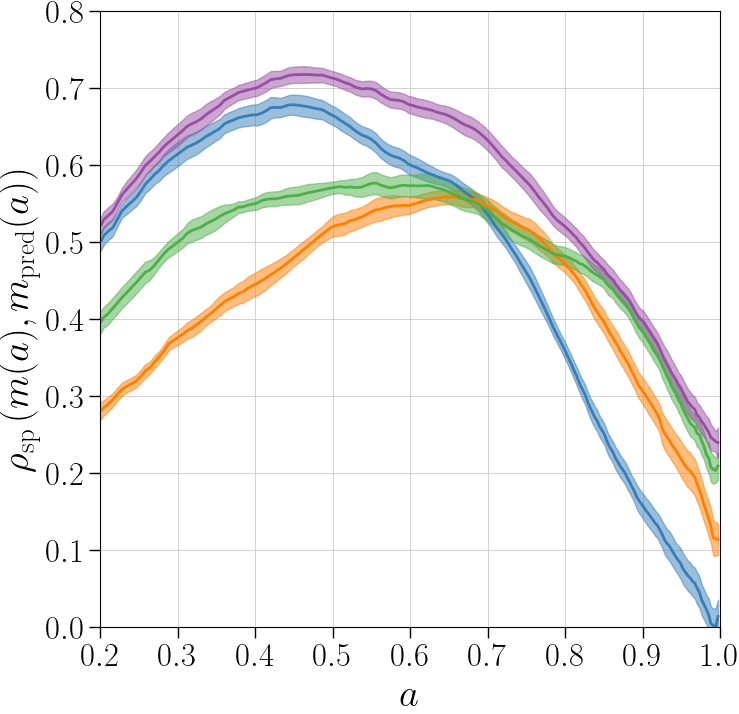}
    \end{subfigure} 
    \quad
    \begin{subfigure}
        \centering
        \includegraphics[width=\columnwidth]{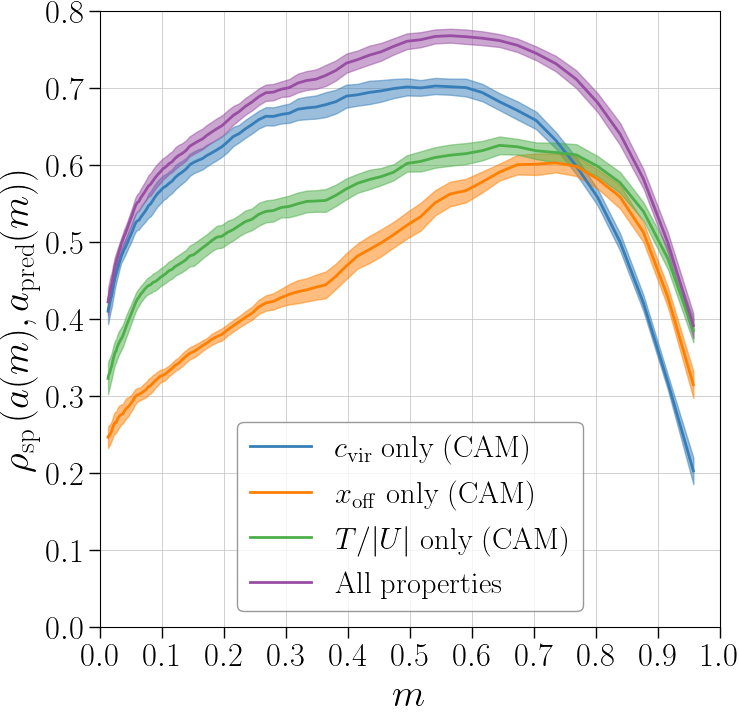}
    \end{subfigure}
    \caption{
        \textbf{MultiCAM predictions of full MAHs based on $z=0$ halo properties.} Here we use the same models as in \cref{fig:inverse-pred-metrics} to predict full MAHs, parametrized by $m(a)$ (left) and $a(m)$ (right). The left-hand curves are the Spearman correlations between the predicted $m(a)$ and the true $m(a)$ at each scale $a$ of the Bolshoi simulation starting at $a=0.2$ for our test set from the \simsample data set. Similarly, the right-hand curves are the Spearman correlations between the predicted $a(m)$ and the true $a(m)$ on uniformly spaced log bins on $m$ between $0$ and $1$. The bands correspond to 68\% confidence intervals, estimated by jackknife resampling. See \cref{sec:pred-mah-with-present} for additional discussion on this figure.
    }
    \label{fig:predict-full-mah}
\end{figure*}

In \cref{fig:predict-full-mah} we use \multicam with the same models as in \cref{fig:inverse-pred-metrics} to predict the full MAH either with the $m(a)$ parametrization (left) or the $a(m)$ parametrization (right). We use the same scale and mass bins for prediction as in \cref{fig:mah-correlations-with-properties} which consist of the Bolshoi simulation cadences for $m(a)$ (left) and uniformly spaced log bins on $m$ between 0 and 1 for $a(m)$ (right). For each parametrization, we show the Spearman correlation coefficient between predicted MAH and true MAH of our test set from the \simsample data set.

\multicam with all properties (purple diamonds) produces strictly more accurate predictions for the mass-accretion histories of haloes than any individual property. It leverages properties like $c_{\rm vir}$ to maintain high accuracy ($\rho_{\rm spearman}\approx 0.75$) at early times and switches to later-time properties like $T/|U|$ and $x_{\rm off}$ to maintain that accuracy after $c_{\rm vir}$ ceases to be a good tracer of growth. 

Overall, the curves follow the same trends as in \cref{fig:mah-correlations-with-properties}. For example, $\xoff$ and $\ToverU$ are better are predicting late history than $\cvir$. $\ToverU$ has higher correlation than $\xoff$ throughout. This is expected since higher Spearman correlation corresponds to higher prediction power. For the same reason as in \cref{fig:mah-correlations-with-properties}, all models predictive power drop at very early and very late times.

Finally, as discussed in \cref{sec:multicam-general}, \multicam reduces to CAM in the case of connecting a single feature with a single target variable. This means that the \multicam predicted correlations for the models using a single halo property as a feature in \cref{fig:inverse-pred-metrics} and \cref{fig:predict-full-mah} are equivalent to the CAM predictions for the corresponding MAH summary.

%%%%%%%%%%%%%%%%%%%%%%%%%%%%%%%%
\section{Discussion} \label{sec:discussion}
%%%%%%%%%%%%%%%%%%%%%%%%%%%%%%%%

% 1. Correlations between halo properties and MAH
In this work, we have studied the correlations between a halo's present-day properties and multiple intermediate epochs of their MAH. In particular, we investigated the time and mass scales at which different halo properties correlate most strongly with the MAH (see \cref{fig:mah-correlations-with-properties} and \cref{tab:optimal-corrs}). We find a significant non-zero correlation between all the halo properties we studied and its formation history for most time and mass scales, with most halo properties, including concentration, achieving their strongest correlation with the MAH at intermediate time and mass scales. 
This is in disagreement with the findings in \citet{wong2012dark}, where the authors find that correlation between concentration and the MAH was strongest when the halo had accumulated only $20\%$ of its mass for a relaxed halo sample. However, we see a high level of agreement both quantitatively and qualitatively for the correlation between concentration and MAH with \cite{wang2020concentrations} and \cite{tae2023splashback}, who both use the same halo finder (\rockstar) as our work. We thus hypothesize that the disagreement with \citet{wong2012dark} is due to differences in halo finder and halo sample, but leave confirmation of this for future work.

% 1.5 MAH autocorrelations
We also studied the autocorrelations between different epochs of mass growth.  \cref{fig:correlation-matrix-am} shows that a sparser representation of the MAH can provide a similar amount of predictive information to model galaxy or halo properties. This conclusion is similar to the one reached in \cite{wong2012dark}, where their principal component analysis of MAHs suggested that only a few principal components explained the majority of the scatter in the MAHs. Physically, this indicates that longer time-scales of mass accretion likely set halo properties.

% 2. Adding to literature on galaxy-halo connection with ML and non-ML methods.
Our model and subsequent analysis adds to a growing body of literature that models the connections between galaxies, dark matter haloes, and their MAHs \citep{wechsler2018connection}.  Such models have ranged from one-to-one mappings of properties in the form of abundance matching \citep{kravtsov2004dark}, to complex machine-learning approaches \citep[e.g.][]{machado2021,hausen2023revealing,horowitz2022differentiable,stiskalek2022,de2023machine}. We provide a generalization of CAM and quantify its ability to connect halo properties with their full MAH. 
Other recent models enable connections between more details of a halo's full MAH with corresponding halo or galaxy properties. For example, \citet{jespersen2022mangrove} builds a graph neural network that directly uses the full dark matter merger tree of a halo to accurately emulate galaxy properties and their scatter. They find that using the full formation history always outperforms predictions compared to only using the $z=0$ halo properties and a traditional abundance matching approach, which is consistent with our conclusions in \cref{fig:forward-pred-metrics} and \cref{fig:inverse-pred-metrics}. 
As another example, \citet{lucie2022insights} uses gradient-boosted-tree algorithms to predict the final mass profiles of cluster-sized haloes based on the initial density field and the MAH. Their model is able to identify time-scales in the MAHs that are most predictive of the final mass profiles. Even though ML approaches such as \citet{jespersen2022mangrove} are inherently non-linear and can achieve high prediction accuracy, MultiCAM allows for easily determining which subset of features contribute toward prediction power and is simpler to invert.

% 3. Future work: include merger information and why we need it.
As demonstrated in \cite{wang2020concentrations} and \cite{rey2019mah}, one major source of scatter in the concentration mass relation comes from mergers, and the scatter depends on fine grained details of these mergers. However, \cref{fig:mah-correlations-with-properties} shows that the last dynamical time of the halo is not providing much predictive information. This suggests that \multicam is not able to successfully extract the relevant merger and non-smooth information from the MAH features given. 
In addition, we attempted to capture merger information by incorporating gradient features of MAH in \multicam's prediction. We found that the prediction performance of \multicam remained the same when adding these additional features across all halo properties. We therefore plan to explicitly incorporate major merger information from merger trees in future development and studies with \multicam.

% 4. Other potential future applications including: predicting baryonic properties, building emulators, using non-linear methods.
Additional future applications of our method include (1) applying \multicam to connecting DM halo accretion histories to baryonic properties in the context of hydrodynamical simulations such as the \textsc{TheThreeHundred} project \citep{haggar2021three}, (2) using \multicam to build fast emulators that paste small-scale properties into cheaply generated ensembles of accurate mock halo catalogues \citep[e.g.][]{tassev2013cola,feng2016fastpm}, or parametric models of MAHs \citep[e.g.][]{hearin2021differentiable}, (3) exploring other extensions of \multicam that incorporate more advanced non-linear methods, such as neural networks, that could provide higher predictive accuracy, and (4) applying \multicam as an empirical method where we can constrain the internal covariance matrix of the model with observational data.

% 5. MutiCAM connects with dynamical state
Finally, previous work indicates that the MAHs closely connect to proxies for the dynamical state of galaxies, galaxy clusters, and their host haloes \citep[e.g.][]{hetznecker2006evolution,gouin2021shape}. An improved understanding of this connection can better inform the interpretation of measurements of galaxy and galaxy cluster properties \citep[e.g.][]{ludlow2012dynamical,mantz2015cosmology,ludlow2016mass}.
The flexibility of \multicam provides a simple and interpretable framework to explore various measures of the dynamical state of DM haloes, galaxies, or galaxy clusters and to see how their dynamical state connects with their structural properties and accretion history.
% Connecting with specific indicators
Specifically, \multicam provides a framework to study the predictive power of any combination of galaxy or halo properties on the MAH. Such studies could enable optimal combinations of properties that strongly correlate with merger information or other indicators of dynamical state. Thus, \multicam complements approaches to classifying the dynamical state of haloes or galaxies similar to the ones proposed in works such as \citet{deluca2021three} and \citet{valles2023choice}, which attempt to construct tracers of halo relaxedness from multiple halo properties. 

%%%%%%%%%%%%%%%%%%%%%%%%%%%%%%%%%%%%%%%%%%%%%%%%%%%%%%%%
\section{Conclusion} \label{sec:conclusions}
%%%%%%%%%%%%%%%%%%%%%%%%%%%%%%%%%%%%%%%%%%%%%%%%%%%%%%%%

In this study, we present \multicam, a generalization of traditional abundance matching algorithms. \multicam connects halo and galaxy properties with their MAHs. As a case study, we apply \multicam to connect the present-day properties of dark matter haloes with their full MAHs using the Bolshoi dark matter-only cosmological simulation.

{\it Our key result is that we can use the entire MAH with \multicam to significantly outperform CAM in such connections.}  Our \multicam models are particularly successful in connecting the entire MAH with halo properties often used to trace MAH, such as $\cvir$, $\ToverU$, and $\xoff$. For other halo properties considered (e.g. $\spin$), \multicam performs at least as well as CAM. See \cref{fig:forward-pred-metrics} and \cref{fig:inverse-pred-metrics} for relevant figures. 

Our other main results are the following:

\begin{enumerate}
    \item There is a significant autocorrelation in dark matter haloes' MAH. We find that values of normalized peak masses $m(a)$ are strongly correlated with one another for small changes in $a$. This indicates that a subset, or a sparser representation of MAH, might be sufficient to model some galaxy or halo properties with comparable information content. For more details, see \cref{fig:correlation-matrix-am}. 
    
    \item The entire formation history of a halo leaves imprints on present-day properties. We find that all the properties in our subset of present-day halo properties have significant non-zero correlations with their MAH between $z\approx 4$ and $z=0$. See \cref{fig:mah-correlations-with-properties}.
    
    \item We find that \multicam applied to the \diffmah smooth parametrization of MAH \citep{hearin2021differentiable} performs comparably with \multicam applied on the full MAH for halo properties known to be strongly correlated with late-time merger events such as $\xoff$ and $\spin$. This suggests that \multicam is not able to fully capture merger information in detail, which we leave for future work. See \cref{fig:forward-pred-metrics} for more details.
    
    \item We show how a simple extension of \multicam based on conditional Gaussian sampling is able to simultaneously sample multiple halo properties based on the MAH and {\it capture the true correlation between properties}. See \cref{fig:3panel-triangle} and \cref{fig:full-triangle}.
    
    \item Finally, we apply \multicam to the inverse problem of predicting the MAH of a halo from its present-day properties. We show that $\cvir$ is better at predicting the early formation history of a halo, and $\ToverU$ and $\xoff$ are better at predicting the late-time formation history. \multicam enables simultaneous use of all halo properties for MAH prediction, which outperforms predictions from any individual property. See \cref{fig:inverse-pred-metrics} and \cref{fig:predict-full-mah}. 
\end{enumerate}

%%%%%%%%%%%%%%%%%%%%%%%%%%%%%%%%%%%%%%%%%%
\section*{Data Availability}
%%%%%%%%%%%%%%%%%%%%%%%%%%%%%%%%%%%%%%%%%%
 
The software to reproduce all results in this work is publicly available in Zenodo, at \url{https://doi.org/10.5281/zenodo.7637864}. Our software is also available at the following public github repository: \url{https://github.com/ismael-mendoza/multicam}. 

The data from the Bolshoi dark matter halo catalogue is publicly available in the CosmoSim data base, at \url{https://doi.org/10.17876/cosmosim/bolshoi}.

%%%%%%%%%%%%%%%%%%%%%%%%%%%%%%%%%%%%%%%%%%
\section*{Acknowledgements}
%%%%%%%%%%%%%%%%%%%%%%%%%%%%%%%%%%%%%%%%%%

IM and CA acknowledge support from DOE grant DE-SC009193. IM, KW, and CA acknowledge support from the Leinweber foundation at the University of Michigan. IM acknowledges the support of the Special Interest Group on High Performance Computing (SIGHPC) Computational and Data Science Fellowship. IM acknowledges support from the Michigan Institute for Computational Discovery and Engineering (MICDE) Graduate Fellowship. 

This research was supported in part through computational resources and services provided by Advanced Research Computing at the University of Michigan, Ann Arbor. The CosmoSim data base used in this paper is a service by the Leibniz-Institute for Astrophysics Potsdam (AIP). The MultiDark data base was developed in cooperation with the Spanish MultiDark Consolider Project CSD2009-00064. The authors gratefully acknowledge the Gauss Centre for Supercomputing e.V. (www.gauss-centre.eu) and the Partnership for Advanced Supercomputing in Europe (PRACE, www.prace-ri.eu) for funding the MultiDark simulation project by providing computing time on the GCS Supercomputer SuperMUC at Leibniz Supercomputing Centre (LRZ, www.lrz.de). The Bolshoi simulations have been performed within the Bolshoi project of the University of California High-Performance AstroComputing Center (UC-HiPACC) and were run at the NASA Ames Research Center.

We acknowledge the use of the \texttt{scikit-learn} software for linear regression models and quantile transformers \citep{scikit-learn}. We also acknowledge the use of \texttt{numpy} \citep{numpy2020}, \texttt{scipy} \citep{scipy2020}, \texttt{colossus} \citep{diemer2018colossus}, \texttt{astropy} \citep{astropy:2013,astropy:2018,astropy2022}, \texttt{matplotlib} \citep{matplotlib2007}, \texttt{corner} \citep{corner2016}, and \texttt{lmfit} \citep{newville2023lmfit}. 

We thank Andrew Hearin and Daisuke Nagai for feedback on early results of our model and analysis.

%%%%%%%%%%%%%%%%%%%% REFERENCES %%%%%%%%%%%%%%%%%%

\bibliographystyle{mnras}
\bibliography{main}

\begin{thebibliography}{}
\makeatletter
\relax
\def\mn@urlcharsother{\let\do\@makeother \do\$\do\&\do\#\do\^\do\_\do\%\do\~}
\def\mn@doi{\begingroup\mn@urlcharsother \@ifnextchar [ {\mn@doi@}
  {\mn@doi@[]}}
\def\mn@doi@[#1]#2{\def\@tempa{#1}\ifx\@tempa\@empty \href
  {http://dx.doi.org/#2} {doi:#2}\else \href {http://dx.doi.org/#2} {#1}\fi
  \endgroup}
\def\mn@eprint#1#2{\mn@eprint@#1:#2::\@nil}
\def\mn@eprint@arXiv#1{\href {http://arxiv.org/abs/#1} {{\tt arXiv:#1}}}
\def\mn@eprint@dblp#1{\href {http://dblp.uni-trier.de/rec/bibtex/#1.xml}
  {dblp:#1}}
\def\mn@eprint@#1:#2:#3:#4\@nil{\def\@tempa {#1}\def\@tempb {#2}\def\@tempc
  {#3}\ifx \@tempc \@empty \let \@tempc \@tempb \let \@tempb \@tempa \fi \ifx
  \@tempb \@empty \def\@tempb {arXiv}\fi \@ifundefined
  {mn@eprint@\@tempb}{\@tempb:\@tempc}{\expandafter \expandafter \csname
  mn@eprint@\@tempb\endcsname \expandafter{\@tempc}}}

\bibitem[\protect\citeauthoryear{{Allgood}, {Flores}, {Primack}, {Kravtsov},
  {Wechsler}, {Faltenbacher}  \& {Bullock}}{{Allgood}
  et~al.}{2006}]{allgood2006shape}
{Allgood} B.,  {Flores} R.~A.,  {Primack} J.~R.,  {Kravtsov} A.~V.,  {Wechsler}
  R.~H.,  {Faltenbacher} A.,   {Bullock} J.~S.,  2006, \mn@doi [\mnras]
  {10.1111/j.1365-2966.2006.10094.x}, \href
  {https://ui.adsabs.harvard.edu/abs/2006MNRAS.367.1781A} {367, 1781}

\bibitem[\protect\citeauthoryear{{Astropy Collaboration}}{{Astropy
  Collaboration}}{2013}]{astropy:2013}
{Astropy Collaboration} 2013, \mn@doi [\aap] {10.1051/0004-6361/201322068},
  \href {https://ui.adsabs.harvard.edu/abs/2013A&A...558A..33A} {558, A33}

\bibitem[\protect\citeauthoryear{{Astropy Collaboration}}{{Astropy
  Collaboration}}{2018}]{astropy:2018}
{Astropy Collaboration} 2018, \mn@doi [\aj] {10.3847/1538-3881/aabc4f}, \href
  {https://ui.adsabs.harvard.edu/abs/2018AJ....156..123A} {156, 123}

\bibitem[\protect\citeauthoryear{{Astropy Collaboration}}{{Astropy
  Collaboration}}{2022}]{astropy2022}
{Astropy Collaboration} 2022, \mn@doi [\apj] {10.3847/1538-4357/ac7c74}, \href
  {https://ui.adsabs.harvard.edu/abs/2022ApJ...935..167A} {935, 167}

\bibitem[\protect\citeauthoryear{{Behroozi}, {Wechsler}  \& {Wu}}{{Behroozi}
  et~al.}{2013a}]{behroozi2012rockstar}
{Behroozi} P.~S.,  {Wechsler} R.~H.,   {Wu} H.-Y.,  2013a, \mn@doi [\apj]
  {10.1088/0004-637X/762/2/109}, \href
  {https://ui.adsabs.harvard.edu/abs/2013ApJ...762..109B} {762, 109}

\bibitem[\protect\citeauthoryear{{Behroozi}, {Wechsler}, {Wu}, {Busha},
  {Klypin}  \& {Primack}}{{Behroozi}
  et~al.}{2013b}]{behroozi2012gravitationally}
{Behroozi} P.~S.,  {Wechsler} R.~H.,  {Wu} H.-Y.,  {Busha} M.~T.,  {Klypin}
  A.~A.,   {Primack} J.~R.,  2013b, \mn@doi [\apj]
  {10.1088/0004-637X/763/1/18}, \href
  {https://ui.adsabs.harvard.edu/abs/2013ApJ...763...18B} {763, 18}

\bibitem[\protect\citeauthoryear{Blumenthal, Faber, Primack  \&
  Rees}{Blumenthal et~al.}{1984}]{blumenthal1984formation}
Blumenthal G.~R.,  Faber S.,  Primack J.~R.,   Rees M.~J.,  1984, Nature, 311,
  517

\bibitem[\protect\citeauthoryear{{Bryan} \& {Norman}}{{Bryan} \&
  {Norman}}{1998}]{bryan1998xrayclusters}
{Bryan} G.~L.,  {Norman} M.~L.,  1998, \mn@doi [\apj] {10.1086/305262}, \href
  {https://ui.adsabs.harvard.edu/abs/1998ApJ...495...80B} {495, 80}

\bibitem[\protect\citeauthoryear{{Chen}, {Avestruz}, {Kravtsov}, {Lau}  \&
  {Nagai}}{{Chen} et~al.}{2019}]{chen19}
{Chen} H.,  {Avestruz} C.,  {Kravtsov} A.~V.,  {Lau} E.~T.,   {Nagai} D.,
  2019, \mn@doi [\mnras] {10.1093/mnras/stz2776}, \href
  {https://ui.adsabs.harvard.edu/abs/2019MNRAS.490.2380C} {490, 2380}

\bibitem[\protect\citeauthoryear{Chen, Mo, Li, Wang, Yang, Zhang  \& Wang}{Chen
  et~al.}{2020}]{chen2020relating}
Chen Y.,  Mo H.,  Li C.,  Wang H.,  Yang X.,  Zhang Y.,   Wang K.,  2020, ApJ,
  899, 81

\bibitem[\protect\citeauthoryear{Cui, Power, Borgani, Knebe, Lewis, Murante  \&
  Poole}{Cui et~al.}{2017}]{cui2017dynamical}
Cui W.,  Power C.,  Borgani S.,  Knebe A.,  Lewis G.~F.,  Murante G.,   Poole
  G.~B.,  2017, Monthly Notices of the Royal Astronomical Society, 464, 2502

\bibitem[\protect\citeauthoryear{Dalal, Lithwick  \& Kuhlen}{Dalal
  et~al.}{2010}]{dalal2010origin}
Dalal N.,  Lithwick Y.,   Kuhlen M.,  2010, preprint (\mn@eprint {arXiv}
  {1010.2539})

\bibitem[\protect\citeauthoryear{De~Luca, De~Petris, Yepes, Cui, Knebe  \&
  Rasia}{De~Luca et~al.}{2021}]{deluca2021three}
De~Luca F.,  De~Petris M.,  Yepes G.,  Cui W.,  Knebe A.,   Rasia E.,  2021,
  Monthly Notices of the Royal Astronomical Society, 504, 5383

\bibitem[\protect\citeauthoryear{Devroye}{Devroye}{1986}]{devroye1986sample}
Devroye L.,  1986, in Proceedings of the 18th Conference on Winter Simulation.
  p.~260

\bibitem[\protect\citeauthoryear{Diemand \& Moore}{Diemand \&
  Moore}{2011}]{diemand2011structure}
Diemand J.,  Moore B.,  2011, Adv. Sci. Lett., 4, 297

\bibitem[\protect\citeauthoryear{{Diemer}}{{Diemer}}{2018}]{diemer2018colossus}
{Diemer} B.,  2018, \mn@doi [\apjs] {10.3847/1538-4365/aaee8c}, \href
  {https://ui.adsabs.harvard.edu/abs/2018ApJS..239...35D} {239, 35}

\bibitem[\protect\citeauthoryear{{Diemer} \& {Kravtsov}}{{Diemer} \&
  {Kravtsov}}{2015}]{diemer2015universal}
{Diemer} B.,  {Kravtsov} A.~V.,  2015, \mn@doi [\apj]
  {10.1088/0004-637X/799/1/108}, \href
  {https://ui.adsabs.harvard.edu/abs/2015ApJ...799..108D} {799, 108}

\bibitem[\protect\citeauthoryear{Dunkley et~al.,}{Dunkley
  et~al.}{2009}]{dunkley2009five}
Dunkley J.,  et~al., 2009, ApJS, 180, 306

\bibitem[\protect\citeauthoryear{Feng, Chu, Seljak  \& McDonald}{Feng
  et~al.}{2016}]{feng2016fastpm}
Feng Y.,  Chu M.-Y.,  Seljak U.,   McDonald P.,  2016, Monthly Notices of the
  Royal Astronomical Society, 463, 2273

\bibitem[\protect\citeauthoryear{{Foreman-Mackey}}{{Foreman-Mackey}}{2016}]{corner2016}
{Foreman-Mackey} D.,  2016, \mn@doi [J. Open Source Softw.]
  {10.21105/joss.00024}, \href
  {https://ui.adsabs.harvard.edu/abs/2016JOSS....1...24F} {1, 24}

\bibitem[\protect\citeauthoryear{Freedman}{Freedman}{2009}]{freedman2009statistical}
Freedman D.~A.,  2009, Statistical Models: Theory and Practice.
Cambridge University Press

\bibitem[\protect\citeauthoryear{Frenk \& White}{Frenk \&
  White}{2012}]{frenk2012dark}
Frenk C.~S.,  White S.~D.,  2012, Ann. Phys., 524, 507

\bibitem[\protect\citeauthoryear{{Gao}, {Springel}  \& {White}}{{Gao}
  et~al.}{2005}]{gao2005age}
{Gao} L.,  {Springel} V.,   {White} S. D.~M.,  2005, \mn@doi [\mnras]
  {10.1111/j.1745-3933.2005.00084.x}, \href
  {https://ui.adsabs.harvard.edu/abs/2005MNRAS.363L..66G} {363, L66}

\bibitem[\protect\citeauthoryear{{Gouin}, {Bonnaire}  \& {Aghanim}}{{Gouin}
  et~al.}{2021}]{gouin2021shape}
{Gouin} C.,  {Bonnaire} T.,   {Aghanim} N.,  2021, \mn@doi [\aap]
  {10.1051/0004-6361/202140327}, \href
  {https://ui.adsabs.harvard.edu/abs/2021A&A...651A..56G} {651, A56}

\bibitem[\protect\citeauthoryear{Haggar, Pearce, Gray, Knebe  \& Yepes}{Haggar
  et~al.}{2021}]{haggar2021three}
Haggar R.,  Pearce F.~R.,  Gray M.~E.,  Knebe A.,   Yepes G.,  2021, Monthly
  Notices of the Royal Astronomical Society, 502, 1191

\bibitem[\protect\citeauthoryear{{Harris} et~al.,}{{Harris}
  et~al.}{2020}]{numpy2020}
{Harris} C.~R.,  et~al., 2020, \mn@doi [\nat] {10.1038/s41586-020-2649-2},
  \href {https://ui.adsabs.harvard.edu/abs/2020Natur.585..357H} {585, 357}

\bibitem[\protect\citeauthoryear{{Hausen}, {Robertson}, {Zhu}, {Gnedin},
  {Madau}, {Schneider}, {Villasenor}  \& {Drakos}}{{Hausen}
  et~al.}{2023}]{hausen2023revealing}
{Hausen} R.,  {Robertson} B.~E.,  {Zhu} H.,  {Gnedin} N.~Y.,  {Madau} P.,
  {Schneider} E.~E.,  {Villasenor} B.,   {Drakos} N.~E.,  2023, \mn@doi [\apj]
  {10.3847/1538-4357/acb25c}, \href
  {https://ui.adsabs.harvard.edu/abs/2023ApJ...945..122H} {945, 122}

\bibitem[\protect\citeauthoryear{{Hearin} \& {Watson}}{{Hearin} \&
  {Watson}}{2013}]{hearin2013darkside}
{Hearin} A.~P.,  {Watson} D.~F.,  2013, \mn@doi [\mnras]
  {10.1093/mnras/stt1374}, \href
  {https://ui.adsabs.harvard.edu/abs/2013MNRAS.435.1313H} {435, 1313}

\bibitem[\protect\citeauthoryear{Hearin, Watson, Becker, Reyes, Berlind  \&
  Zentner}{Hearin et~al.}{2014}]{hearin2014dark}
Hearin A.~P.,  Watson D.~F.,  Becker M.~R.,  Reyes R.,  Berlind A.~A.,
  Zentner A.~R.,  2014, Monthly Notices of the Royal Astronomical Society, 444,
  729

\bibitem[\protect\citeauthoryear{Hearin, Zentner, van~den Bosch, Campbell  \&
  Tollerud}{Hearin et~al.}{2016}]{hearin2016introducing}
Hearin A.~P.,  Zentner A.~R.,  van~den Bosch F.~C.,  Campbell D.,   Tollerud
  E.,  2016, Monthly Notices of the Royal Astronomical Society, 460, 2552

\bibitem[\protect\citeauthoryear{{Hearin}, {Chaves-Montero}, {Becker}  \&
  {Alarcon}}{{Hearin} et~al.}{2021}]{hearin2021differentiable}
{Hearin} A.~P.,  {Chaves-Montero} J.,  {Becker} M.~R.,   {Alarcon} A.,  2021,
  \mn@doi [Open J. Astrophys.] {10.21105/astro.2105.05859}, \href
  {https://ui.adsabs.harvard.edu/abs/2021OJAp....4E...7H} {4, 7}

\bibitem[\protect\citeauthoryear{Hetznecker \& Burkert}{Hetznecker \&
  Burkert}{2006}]{hetznecker2006evolution}
Hetznecker H.,  Burkert A.,  2006, Monthly Notices of the Royal Astronomical
  Society, 370, 1905

\bibitem[\protect\citeauthoryear{{Horowitz}, {Hahn}, {Lanusse}, {Modi}  \&
  {Ferraro}}{{Horowitz} et~al.}{2022}]{horowitz2022differentiable}
{Horowitz} B.,  {Hahn} C.,  {Lanusse} F.,  {Modi} C.,   {Ferraro} S.,  2022,
  preprint (\mn@eprint {arXiv} {2211.03852})

\bibitem[\protect\citeauthoryear{{Hunter}}{{Hunter}}{2007}]{matplotlib2007}
{Hunter} J.~D.,  2007, \mn@doi [Comput. Sci. Eng.] {10.1109/MCSE.2007.55},
  \href {https://ui.adsabs.harvard.edu/abs/2007CSE.....9...90H} {9, 90}

\bibitem[\protect\citeauthoryear{{Jespersen}, {Cranmer}, {Melchior}, {Ho},
  {Somerville}  \& {Gabrielpillai}}{{Jespersen}
  et~al.}{2022}]{jespersen2022mangrove}
{Jespersen} C.~K.,  {Cranmer} M.,  {Melchior} P.,  {Ho} S.,  {Somerville}
  R.~S.,   {Gabrielpillai} A.,  2022, \mn@doi [\apj]
  {10.3847/1538-4357/ac9b18}, \href
  {https://ui.adsabs.harvard.edu/abs/2022ApJ...941....7J} {941, 7}

\bibitem[\protect\citeauthoryear{Klypin, Trujillo-Gomez  \& Primack}{Klypin
  et~al.}{2011}]{klypin2011dark}
Klypin A.~A.,  Trujillo-Gomez S.,   Primack J.,  2011, ApJ, 740, 102

\bibitem[\protect\citeauthoryear{Klypin, Yepes, Gottl{\"o}ber, Prada  \&
  Hess}{Klypin et~al.}{2016}]{klypin2016multidark}
Klypin A.,  Yepes G.,  Gottl{\"o}ber S.,  Prada F.,   Hess S.,  2016, Monthly
  Notices of the Royal Astronomical Society, 457, 4340

\bibitem[\protect\citeauthoryear{{Knebe} et~al.,}{{Knebe}
  et~al.}{2011}]{knebe2011halosgonemad}
{Knebe} A.,  et~al., 2011, \mn@doi [\mnras] {10.1111/j.1365-2966.2011.18858.x},
  \href {https://ui.adsabs.harvard.edu/abs/2011MNRAS.415.2293K} {415, 2293}

\bibitem[\protect\citeauthoryear{Kravtsov, Klypin  \& Khokhlov}{Kravtsov
  et~al.}{1997}]{kravtsov1997adaptive}
Kravtsov A.~V.,  Klypin A.~A.,   Khokhlov A.~M.,  1997, ApJS, 111, 73

\bibitem[\protect\citeauthoryear{Kravtsov, Berlind, Wechsler, Klypin,
  Gottl{\"o}ber, Allgood  \& Primack}{Kravtsov et~al.}{2004}]{kravtsov2004dark}
Kravtsov A.~V.,  Berlind A.~A.,  Wechsler R.~H.,  Klypin A.~A.,  Gottl{\"o}ber
  S.,  Allgood B.,   Primack J.~R.,  2004, ApJ, 609, 35

\bibitem[\protect\citeauthoryear{Lau, Hearin, Nagai  \& Cappelluti}{Lau
  et~al.}{2021}]{lau2021correlations}
Lau E.~T.,  Hearin A.~P.,  Nagai D.,   Cappelluti N.,  2021, Monthly Notices of
  the Royal Astronomical Society, 500, 1029

\bibitem[\protect\citeauthoryear{{Lucie-Smith}, {Adhikari}  \&
  {Wechsler}}{{Lucie-Smith} et~al.}{2022}]{lucie2022insights}
{Lucie-Smith} L.,  {Adhikari} S.,   {Wechsler} R.~H.,  2022, \mn@doi [\mnras]
  {10.1093/mnras/stac1833}, \href
  {https://ui.adsabs.harvard.edu/abs/2022MNRAS.515.2164L} {515, 2164}

\bibitem[\protect\citeauthoryear{Ludlow, Navarro, Li, Angulo, Boylan-Kolchin
  \& Bett}{Ludlow et~al.}{2012}]{ludlow2012dynamical}
Ludlow A.~D.,  Navarro J.~F.,  Li M.,  Angulo R.~E.,  Boylan-Kolchin M.,   Bett
  P.~E.,  2012, Monthly Notices of the Royal Astronomical Society, 427, 1322

\bibitem[\protect\citeauthoryear{{Ludlow} et~al.,}{{Ludlow}
  et~al.}{2013}]{ludlow2013massprofileL}
{Ludlow} A.~D.,  et~al., 2013, \mn@doi [\mnras] {10.1093/mnras/stt526}, \href
  {https://ui.adsabs.harvard.edu/abs/2013MNRAS.432.1103L} {432, 1103}

\bibitem[\protect\citeauthoryear{Ludlow, Bose, Angulo, Wang, Hellwing, Navarro,
  Cole  \& Frenk}{Ludlow et~al.}{2016}]{ludlow2016mass}
Ludlow A.~D.,  Bose S.,  Angulo R.~E.,  Wang L.,  Hellwing W.~A.,  Navarro
  J.~F.,  Cole S.,   Frenk C.~S.,  2016, Monthly Notices of the Royal
  Astronomical Society, 460, 1214

\bibitem[\protect\citeauthoryear{Maccio, Dutton, Van Den~Bosch, Moore, Potter
  \& Stadel}{Maccio et~al.}{2007}]{maccio2007concentration}
Maccio A.~V.,  Dutton A.~A.,  Van Den~Bosch F.~C.,  Moore B.,  Potter D.,
  Stadel J.,  2007, Monthly Notices of the Royal Astronomical Society, 378, 55

\bibitem[\protect\citeauthoryear{{Machado Poletti Valle}, {Avestruz}, {Barnes},
  {Farahi}, {Lau}  \& {Nagai}}{{Machado Poletti Valle}
  et~al.}{2021}]{machado2021}
{Machado Poletti Valle} L.~F.,  {Avestruz} C.,  {Barnes} D.~J.,  {Farahi} A.,
  {Lau} E.~T.,   {Nagai} D.,  2021, \mn@doi [\mnras] {10.1093/mnras/stab2252},
  \href {https://ui.adsabs.harvard.edu/abs/2021MNRAS.507.1468M} {507, 1468}

\bibitem[\protect\citeauthoryear{{Mansfield} \& {Avestruz}}{{Mansfield} \&
  {Avestruz}}{2021}]{mansfield2021biased}
{Mansfield} P.,  {Avestruz} C.,  2021, \mn@doi [\mnras]
  {10.1093/mnras/staa3388}, \href
  {https://ui.adsabs.harvard.edu/abs/2021MNRAS.500.3309M} {500, 3309}

\bibitem[\protect\citeauthoryear{Mantz, Allen, Morris, Schmidt, von~der Linden
  \& Urban}{Mantz et~al.}{2015}]{mantz2015cosmology}
Mantz A.~B.,  Allen S.~W.,  Morris R.~G.,  Schmidt R.~W.,  von~der Linden A.,
  Urban O.,  2015, Monthly Notices of the Royal Astronomical Society, 449, 199

\bibitem[\protect\citeauthoryear{{Mo}, {van den Bosch}  \& {White}}{{Mo}
  et~al.}{2010}]{mo2010thebook}
{Mo} H.,  {van den Bosch} F.~C.,   {White} S.,  2010, {Galaxy Formation and
  Evolution}.
Cambridge University Press

\bibitem[\protect\citeauthoryear{Neto et~al.,}{Neto
  et~al.}{2007}]{neto2007statistics}
Neto A.~F.,  et~al., 2007, Monthly Notices of the Royal Astronomical Society,
  381, 1450

\bibitem[\protect\citeauthoryear{Newville et~al.,}{Newville
  et~al.}{2023}]{newville2023lmfit}
Newville M.,  et~al., 2023, lmfit/lmfit-py: 1.2.1,
  \mn@doi{10.5281/zenodo.7887568}

\bibitem[\protect\citeauthoryear{Pedregosa et~al.,}{Pedregosa
  et~al.}{2011}]{scikit-learn}
Pedregosa F.,  et~al., 2011, J. Mach. Learn. Res., 12, 2825

\bibitem[\protect\citeauthoryear{Power, Knebe  \& Knollmann}{Power
  et~al.}{2012}]{power2012dynamical}
Power C.,  Knebe A.,   Knollmann S.~R.,  2012, Monthly Notices of the Royal
  Astronomical Society, 419, 1576

\bibitem[\protect\citeauthoryear{{Rey}, {Pontzen}  \& {Saintonge}}{{Rey}
  et~al.}{2019}]{rey2019mah}
{Rey} M.~P.,  {Pontzen} A.,   {Saintonge} A.,  2019, \mn@doi [\mnras]
  {10.1093/mnras/stz552}, \href
  {https://ui.adsabs.harvard.edu/abs/2019MNRAS.485.1906R} {485, 1906}

\bibitem[\protect\citeauthoryear{{Rodr{\'\i}guez-Puebla}, {Behroozi},
  {Primack}, {Klypin}, {Lee}  \& {Hellinger}}{{Rodr{\'\i}guez-Puebla}
  et~al.}{2016}]{rodriguez_peubla2016demographics}
{Rodr{\'\i}guez-Puebla} A.,  {Behroozi} P.,  {Primack} J.,  {Klypin} A.,  {Lee}
  C.,   {Hellinger} D.,  2016, \mn@doi [\mnras] {10.1093/mnras/stw1705}, \href
  {https://ui.adsabs.harvard.edu/abs/2016MNRAS.462..893R} {462, 893}

\bibitem[\protect\citeauthoryear{{Savitzky} \& {Golay}}{{Savitzky} \&
  {Golay}}{1964}]{savitzky1964smoothing}
{Savitzky} A.,  {Golay} M.~J.~E.,  1964, \mn@doi [Anal. Chem.]
  {10.1021/ac60214a047}, \href
  {https://ui.adsabs.harvard.edu/abs/1964AnaCh..36.1627S} {36, 1627}

\bibitem[\protect\citeauthoryear{{Sheth}, {Mo}  \& {Tormen}}{{Sheth}
  et~al.}{2001}]{sheth2001ellipsoidal}
{Sheth} R.~K.,  {Mo} H.~J.,   {Tormen} G.,  2001, \mn@doi [\mnras]
  {10.1046/j.1365-8711.2001.04006.x}, \href
  {https://ui.adsabs.harvard.edu/abs/2001MNRAS.323....1S} {323, 1}

\bibitem[\protect\citeauthoryear{{Shin} \& {Diemer}}{{Shin} \&
  {Diemer}}{2023}]{tae2023splashback}
{Shin} T.-h.,  {Diemer} B.,  2023, \mn@doi [\mnras] {10.1093/mnras/stad860},
  \href {https://ui.adsabs.harvard.edu/abs/2023MNRAS.521.5570S} {521, 5570}

\bibitem[\protect\citeauthoryear{{Stiskalek}, {Bartlett}, {Desmond}  \&
  {Anbajagane}}{{Stiskalek} et~al.}{2022}]{stiskalek2022}
{Stiskalek} R.,  {Bartlett} D.~J.,  {Desmond} H.,   {Anbajagane} D.,  2022,
  \mn@doi [\mnras] {10.1093/mnras/stac1609}, \href
  {https://ui.adsabs.harvard.edu/abs/2022MNRAS.514.4026S} {514, 4026}

\bibitem[\protect\citeauthoryear{Tassev, Zaldarriaga  \& Eisenstein}{Tassev
  et~al.}{2013}]{tassev2013cola}
Tassev S.,  Zaldarriaga M.,   Eisenstein D.~J.,  2013, J. Cosmol. Astropart.
  Phys., 2013, 036

\bibitem[\protect\citeauthoryear{Tormen, Bouchet  \& White}{Tormen
  et~al.}{1997}]{tormen1997structure}
Tormen G.,  Bouchet F.~R.,   White S.~D.,  1997, Monthly Notices of the Royal
  Astronomical Society, 286, 865

\bibitem[\protect\citeauthoryear{{Tucci}, {Montero-Dorta}, {Abramo},
  {Sato-Polito}  \& {Artale}}{{Tucci} et~al.}{2021}]{tucci2021spinbias}
{Tucci} B.,  {Montero-Dorta} A.~D.,  {Abramo} L.~R.,  {Sato-Polito} G.,
  {Artale} M.~C.,  2021, \mn@doi [\mnras] {10.1093/mnras/staa3319}, \href
  {https://ui.adsabs.harvard.edu/abs/2021MNRAS.500.2777T} {500, 2777}

\bibitem[\protect\citeauthoryear{{Vall{\'e}s-P{\'e}rez}, {Planelles},
  {Monllor-Berbegal}  \& {Quilis}}{{Vall{\'e}s-P{\'e}rez}
  et~al.}{2023}]{valles2023choice}
{Vall{\'e}s-P{\'e}rez} D.,  {Planelles} S.,  {Monllor-Berbegal} {\'O}.,
  {Quilis} V.,  2023, \mn@doi [\mnras] {10.1093/mnras/stad059}, \href
  {https://ui.adsabs.harvard.edu/abs/2023MNRAS.519.6111V} {519, 6111}

\bibitem[\protect\citeauthoryear{{Virtanen} et~al.,}{{Virtanen}
  et~al.}{2020}]{scipy2020}
{Virtanen} P.,  et~al., 2020, \mn@doi [Nature Methods]
  {10.1038/s41592-019-0686-2}, \href
  {https://ui.adsabs.harvard.edu/abs/2020NatMe..17..261V} {17, 261}

\bibitem[\protect\citeauthoryear{{Vitvitska}, {Klypin}, {Kravtsov}, {Wechsler},
  {Primack}  \& {Bullock}}{{Vitvitska} et~al.}{2002}]{vitvitska2002spin}
{Vitvitska} M.,  {Klypin} A.~A.,  {Kravtsov} A.~V.,  {Wechsler} R.~H.,
  {Primack} J.~R.,   {Bullock} J.~S.,  2002, \mn@doi [\apj] {10.1086/344361},
  \href {https://ui.adsabs.harvard.edu/abs/2002ApJ...581..799V} {581, 799}

\bibitem[\protect\citeauthoryear{{Wang}, {Mao}, {Zentner}, {Lange}, {van den
  Bosch}  \& {Wechsler}}{{Wang} et~al.}{2020}]{wang2020concentrations}
{Wang} K.,  {Mao} Y.-Y.,  {Zentner} A.~R.,  {Lange} J.~U.,  {van den Bosch}
  F.~C.,   {Wechsler} R.~H.,  2020, \mn@doi [\mnras] {10.1093/mnras/staa2733},
  \href {https://ui.adsabs.harvard.edu/abs/2020MNRAS.498.4450W} {498, 4450}

\bibitem[\protect\citeauthoryear{{Watson} et~al.,}{{Watson}
  et~al.}{2015}]{watson2015predicting}
{Watson} D.~F.,  et~al., 2015, \mn@doi [\mnras] {10.1093/mnras/stu2065}, \href
  {https://ui.adsabs.harvard.edu/abs/2015MNRAS.446..651W} {446, 651}

\bibitem[\protect\citeauthoryear{{Wechsler} \& {Tinker}}{{Wechsler} \&
  {Tinker}}{2018}]{wechsler2018connection}
{Wechsler} R.~H.,  {Tinker} J.~L.,  2018, \mn@doi [\araa]
  {10.1146/annurev-astro-081817-051756}, \href
  {https://ui.adsabs.harvard.edu/abs/2018ARA&A..56..435W} {56, 435}

\bibitem[\protect\citeauthoryear{{Wechsler}, {Bullock}, {Primack}, {Kravtsov}
  \& {Dekel}}{{Wechsler} et~al.}{2002}]{wechsler2002concentrationsfromassembly}
{Wechsler} R.~H.,  {Bullock} J.~S.,  {Primack} J.~R.,  {Kravtsov} A.~V.,
  {Dekel} A.,  2002, \mn@doi [\apj] {10.1086/338765}, \href
  {https://ui.adsabs.harvard.edu/abs/2002ApJ...568...52W} {568, 52}

\bibitem[\protect\citeauthoryear{White \& Rees}{White \&
  Rees}{1978}]{white1978core}
White S.~D.,  Rees M.~J.,  1978, Monthly Notices of the Royal Astronomical
  Society, 183, 341

\bibitem[\protect\citeauthoryear{Wong \& Taylor}{Wong \&
  Taylor}{2012}]{wong2012dark}
Wong A.~W.,  Taylor J.~E.,  2012, ApJ, 757, 102

\bibitem[\protect\citeauthoryear{Zhang, Cui, Wang, Dave  \& DePetris}{Zhang
  et~al.}{2022}]{zhang2022three}
Zhang B.,  Cui W.,  Wang Y.,  Dave R.,   DePetris M.,  2022, Monthly Notices of
  the Royal Astronomical Society, 516, 26

\bibitem[\protect\citeauthoryear{de Andres, Yepes, Sembolini,
  Mart{\'\i}nez-Mu{\~n}oz, Cui, Robledo, Chuang  \& Rasia}{de~Andres
  et~al.}{2023}]{de2023machine}
de Andres D.,  Yepes G.,  Sembolini F.,  Mart{\'\i}nez-Mu{\~n}oz G.,  Cui W.,
  Robledo F.,  Chuang C.-H.,   Rasia E.,  2023, Monthly Notices of the Royal
  Astronomical Society, 518, 111

\makeatother
\end{thebibliography}

%%%%%%%%%%%%%%%%% APPENDICES %%%%%%%%%%%%%%%%%%%%%

\appendix
\crefalias{section}{appendix}

\vspace{-6pt}
%%%%%%%%%%%%%%%%%%%%%%%%%%%%%%%%%%%%%%%%%%%%%%%%%%%%%%%%
\section{Conditional Multivariate Gaussian Sampling Algorithm} \label{app:multivariate-gaussian-algorithm}
%%%%%%%%%%%%%%%%%%%%%%%%%%%%%%%%%%%%%%%%%%%%%%%%%%%%%%%%

Let $\xv \in \mathbb{R}^{m}$ be a vector of features and $\yv \in \mathbb{R}^{\ell}$ be a vector of targets. We also assume we have access to a training data set of $n$ pairs: $\{\left(\vec{x}_{i}, \yv_{i}\right)\}_{i=1}^{n}$.

The conditional multivariate Gaussian sampling approach consists of the following steps: 

\begin{itemize}
	\item Assume that $\vec{x}, \vec{y}$ are jointly Gaussian distributed so that: 
	\begin{equation}
		\begin{bmatrix}
			\yv \\ 
			\xv
		\end{bmatrix} \sim \mathcal{N}(\vec{\mu}, \vec{\Sigma}) = \mathcal{N} \left(
		\begin{bmatrix}
			\vec{\mu}_y \\ 
			\vec{\mu}_{x}
		\end{bmatrix}, 	
		\begin{bmatrix}
		   	\vec{\Sigma}_{yy} & \vec{\Sigma}_{yx} \\
			\vec{\Sigma}_{xy} & \vec{\Sigma}_{xx}
	 	\end{bmatrix}
        \right)
	\end{equation}
	where we separated the mean $\vec{\mu}$ and covariance matrix $\vec{\Sigma}$ into corresponding blocks for each variable.

	\item Empirically compute estimates for the mean $\vec{\mu}$ and the covariance matrix $\vec{\Sigma}$ using the training set $\{\left(x_{i}, y_{i}\right)\}_{i=1}^{n}$. For example: 
	\begin{align}
	    \hat{\mu}_{x,k} &= \frac{1}{n} \sum_{i=1}^{n} x_{ik} \label{eq:mu-x} \\ 
		\hat{\Sigma}_{xx, k\ell} &= \sum_{i=1}^{n} \frac{(x_{ik} - \hat{\mu}_{x,k}) (x_{i\ell} - \hat{\mu}_{x, \ell})}{n-1} \label{eq:sigma-xx}
    \end{align}
	\item Given a feature vector in the testing set $\vec{x}_{0}$ we are interested in obtaining a prediction $\hat{\vec{y}}_{0}$ for the corresponding true target $\vec{y}_{0}$. In order to do this we need to determine the conditional distribution $P(\yv | \xv_{0})$.

	From statistics we know that the conditional distribution of two or more jointly normal distributed variables is also normal, in particular: 
    \vspace{-6.5pt}
    \begin{equation}
		\vec{y} | \xv_{0} \sim \mathcal{N}(\bar{\vec{\mu}} (\xv_{0}), \bar{\vec{\Sigma}}) \label{eq:gaussian-posterior}
	\end{equation}
	where 
	\begin{align}
		\bar{\vec{\mu}} (\xv_{0}) &= \vec{\mu}_{y} + \vec{\Sigma}_{yx} \vec{\Sigma}_{xx}^{-1} (\xv_{0} - \vec{\mu}_{x}) \label{eq:mu_bar} \\
		\bar{\vec{\Sigma}} &= \vec{\Sigma}_{yy} - \vec{\Sigma}_{yx} \vec{\Sigma}_{xx}^{-1} \vec{\Sigma}_{xy} \label{eq:bar-sigma}
	\end{align}
	Note that to calculate this quantities we would replace $\vec{\mu}$ and $\vec{\Sigma}$ with their corresponding empirical estimates $\hat{\vec{\mu}}, \hat{\vec{\Sigma}}$. Importantly $\bar{\vec{\mu}}$ depends on a particular test point $\xv_{0}$ but $\bar{\vec{\Sigma}}$ does not. 
	
	\item Finally, from \cref{eq:gaussian-posterior} we see that there are two natural options for our predictor $\hat{\vec{y}}_{0}$. We could choose to simply make our predictor the mean of the posterior distribution, i.e., setting $\hat{\vec{y}}_{0} = \bar{\vec{\mu}}(\xv_{0})$. This is reasonable since $\bar{\vec{\mu}} (x_{0})$ is the most likely value of $\vec{y}_{0}$ given $\vec{x}_{0}$ (assuming our statistical model is correct). In fact, in \Cref{app:lr-is-special} we show that this approach is equivalent to linear regression.
	
	Another option is to sample from the distribution in \cref{eq:gaussian-posterior} and in this way account for scatter. This is the approach used in \cref{sec:multicam-scatter}, and it allows us to accurately capture correlations between target variables. This latter approach is what we refer to as conditional Gaussian sampling.
\end{itemize}

%%%%%%%%%%%%%%%%%%%%%%%%%%%%%%%%%%%%%%%%%%%%%%%%%%%%%%%%
\section{Linear Regression: Scatter Agnostic Predictions from Multivariate Gaussian} \label{app:lr-is-special}
%%%%%%%%%%%%%%%%%%%%%%%%%%%%%%%%%%%%%%%%%%%%%%%%%%%%%%%%

In this short appendix, we will show that the predictions from linear regression and the conditional multivariate Gaussian sampling algorithm in \cref{sec:statistical-algorithms} are equivalent up to the scatter coming from $\bar{\vec{\Sigma}}$ (equation \ref{eq:bar-sigma}) in the Gaussian approach. In other words, linear regression outputs the mean of the posterior $P(\vec{y} | \vec{x})$ (equation \ref{eq:gaussian-posterior}) outputted by the multi-Gaussian approach.

Let $\vec{X} \in \mathbb{R}^{n \times m}$ be a training data set of features and $\vec{Y} \in \mathbb{R}^{n \times \ell}$ a training set of predictors. The matrix $\vec{X}$ contains $n$ training feature vectors each with dimensionality $m$, and $\vec{Y}$ contains $n$ predictors each of dimensionality $\ell$. Explicitly, the matrices are:
\begin{align*}
	\vec{X} = \begin{bmatrix}
			\xv_{1}^{T} \\ 
			\xv_{2}^{T} \\ 
			\vdots \\ 
			\xv_{n}^{T}
		\end{bmatrix}, 	\\
	\vec{Y} = \begin{bmatrix} 
		\vec{y}^{T}_{1} \\ 
		\vec{y}^{T}_{2} \\ 
		\vdots \\
		\vec{y}^{T}_{n}
	\end{bmatrix}.
\end{align*}
For simplicity, we assume that our data set has been mean-centred so that $\hat{\vec{\mu}}_{x} = 0$ and $\hat{\vec{\mu}}_{y} = 0$ (equation \ref{eq:mu-x}). 

From standard statistical literature \citep[see, e.g.,][]{freedman2009statistical}, we know that given a new data point $\vec{x}_{0} \in \mathbb{R}^{m}$ the linear regression prediction $\hat{\vec{y}}_{0}$ is: 
\begin{equation}
    \hat{\vec{y}}_{0} = \vec{Y}^{T} \vec{X} (\vec{X}^{T} \vec{X})^{-1} \vec{x}_{0}.
    \label{eq:linear-y0-prediction}
\end{equation}
Let us rewrite the matrix $\vec{X}^{T}\vec{X}$. Given that $X_{ij} = x_{ij}$ we have from \cref{eq:sigma-xx} that: 
\begin{equation}
	(X^{T} X)_{ij} = \sum_{k} x_{ki} x_{kj} = (\hat{\Sigma}_{xx})_{ij} (n-1),
\end{equation}
where the last equality holds since we assumed $\hat{\vec{\mu}}_{x} = 0$. Similarly $\vec{Y}^{T} \vec{X} = \hat{\Sigma}_{yx} (n-1)$, so that combining this with \cref{eq:linear-y0-prediction} we get:
\begin{equation}
	\hat{\yv}_{0} = \hat{\Sigma}_{yx} (n-1) \hat{\Sigma}_{xx}^{-1} (n-1)^{-1}  \xv_{0} = \hat{\Sigma}_{yx} \hat{\Sigma}_{xx}^{-1} \xv_{0},
\end{equation}
which is the same as \cref{eq:mu_bar} in the mean-centred case. This proves that linear regression is the same as multivariate Gaussian sampling without scatter.

%%%%%%%%%%%%%%%%%%%%%%%%%%%%%%%%%%%%%%%%%%%%%%%%%%%%%%%%
\section{MultiCAM captures covariances of present-day halo properties} \label{app:triangle-covariances}
%%%%%%%%%%%%%%%%%%%%%%%%%%%%%%%%%%%%%%%%%%%%%%%%%%%%%%%%

\begin{figure*}
    \centering
    \includegraphics[width=\textwidth]{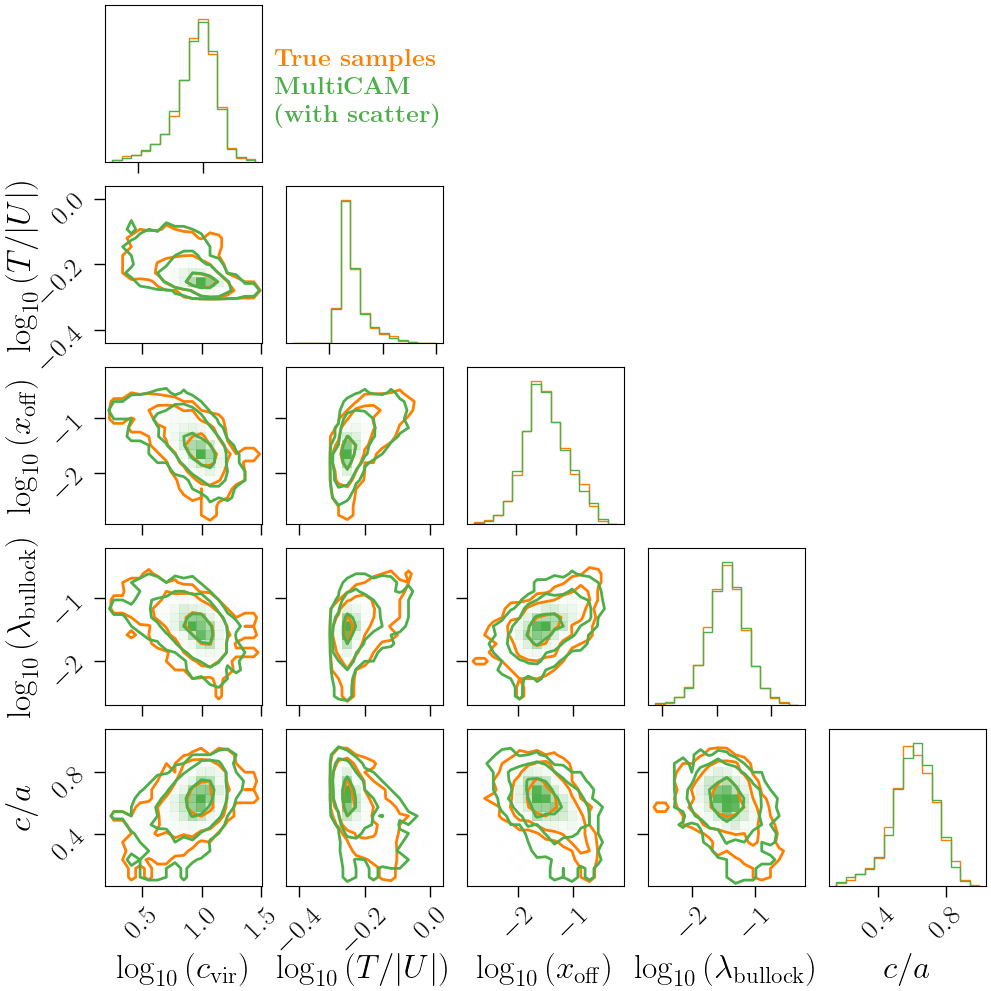}
    \caption{
        \textbf{1D Marginals and 2D Scatter with contours of samples of $z=0$ halo properties from \multicam (with scatter).} We show plots of 1D marginals and of $1\sigma$, $2\sigma$, and $3\sigma$ contours for the 2D histograms of $3000$ samples of $z=0$ halo properties given their full MAH using \multicam (with scatter).
        The orange contours correspond to a sample of $\cvir, \ToverU, \xoff, \spin$, and $c/a$ from $3000$ haloes in our \simsample data set.
        The green contours of each subplot were produced by applying \multicam (with scatter) on the full MAH of each of the $3000$ haloes. See \cref{sec:multicam-scatter} for more discussion on \multicam (with scatter). See \Cref{app:triangle-covariances} for additional discussion of this figure.
    }
\label{fig:full-triangle}
\end{figure*}

In \cref{fig:full-triangle}, we plot 1D marginals and 2D histograms with $1\sigma$, $2\sigma$, and $3\sigma$ contours of our main present-day properties of 3000 haloes from our $\simsample$ data set. The orange contours come from the true values of these halo properties and the green contours are samples from \multicam (with scatter) applied on the full MAH of each of these 3000 haloes. We overall see good agreement between truth (green) and samples from our model (orange) in both the 1D marginal distributions and the 2D scatter contour plots. 

%%%%%%%%%%%%%%%%%%%%%%%%%%%%%%%%%%%%%%%%%%%%%%%%%%

% Don't change these lines
\bsp	% typesetting comment
\label{lastpage}
\end{document}